\documentclass[preprint,12pt]{elsarticle}
\usepackage{hyperref}

\usepackage{graphicx}
\usepackage{epstopdf}
\usepackage{amssymb}
\usepackage{amsthm}
\usepackage{amsmath}
\usepackage{mathtools}
\usepackage{color}
\DeclarePairedDelimiter{\Abs}{\lvert}{\rvert}

\usepackage{soul,xcolor}

\usepackage{subcaption}

\journal{Chaos, Solitons \& Fractals}

\begin{document}

\setstcolor{red}

\begin{frontmatter}


\title{Ordinal Synchronization: Using ordinal patterns to capture interdependencies between time series}


\author[ctb,urjc]{I.~Echegoyen\corref{cor1}}
\author[udg]{V.~Vera-\'Avila}
\author[udg]{R.~Sevilla-Escoboza}
\author[inserm,urjc]{J.H.~Mart\'inez}
\author[ctb,urjc]{J.M.~Buld\'u}
\cortext[cor1]{Corresponding author. \textit{E-mail address: nacho.e.blanco@gmail.com}}

\address[ctb]{Laboratory of Biological Networks, Centre for Biomedical Technology. Universidad Polit\'ecnica de Madrid, Spain}

\address[urjc]{Complex Systems Group \& G.I.S.C., Universidad Rey Juan Carlos, M\'ostoles, Spain}

\address[udg]{Centro Universitario de Los Lagos, Universidad de Guadalajara, Lagos de Moreno, Mexico}

\address[inserm]{INSERM-UM1127, Sorbonne Universit\'e, ICM-H\^opital Piti\'e Salp\^etri\`ere, Paris, France}

\tnotetext[t1]{I. Echegoyen would like to thank the Foundation Tatiana P\'erez de Guzm\'an el Bueno for financially supporting this research. R.S.E. acknowledges support from Consejo Nacional de Ciencia y Tecnolog\'ia call SEP-CONACYT/CB-2016-01, grant number 285909. J.M.B. is supported by Spanish Ministry of Economy and Competitiveness under Projects FIS2013-41057-P and FIS2017-84151-P. J.H.M. acknowledges M. Chavez for his valuable comments.}



\begin{abstract}
We introduce Ordinal Synchronization ($OS$) as a new measure to quantify synchronization between dynamical systems. $OS$ is calculated from the extraction of the ordinal patterns related to two time series, their transformation into $D$-dimensional ordinal vectors and the adequate quantification of their alignment. $OS$  provides a fast and robust-to noise tool to assess synchronization without any implicit assumption about the distribution of data sets nor their dynamical properties, capturing in-phase and anti-phase synchronization. Furthermore, varying the length of the ordinal vectors required to compute $OS$ it is possible to detect synchronization at different time scales. We test the performance of $OS$ with data sets coming from unidirectionally coupled electronic Lorenz oscillators and brain imaging datasets obtained from magnetoencephalographic recordings, comparing the performance of $OS$ with other classical metrics that quantify synchronization between dynamical systems.
\end{abstract}


\end{frontmatter}


Since the seminal work of Huygens about the coordinated motion of two pendulum clocks (refereed to as ``an odd kind of sympathy") \cite{huygens1893}, the study of synchronization in real systems has been one of the major research lines in nonlinear dynamics. From fireflies to neurons, synchronization has been reported in a diversity of social (e.g., human movement or clapping) \cite{neda2000,schmidt1990}, biological (e.g., brain regions or cardiac tissue) \cite{varela2001,agladze2017} and technological systems (e.g., wireless communications or power grids) \cite{romer2001,wang2016}, being in many cases a fundamental process for the functioning of the underlying system. 
However, despite being an ubiquitous phenomenon, the detection and quantification of synchronization can be a difficult task.
The main reasons are the diversity of kinds of synchronization \cite{arenas2008}, the complexity of interaction between dynamical systems \cite{boccaletti2006}, the existence of unavoidable external perturbations \cite{mones2014} or the inability of observing all variables of a real system \cite{rech1990}, just to name a few.

As a consequence, there is not a unique way of quantifying the amount of synchronization in real time series and a series of metrics have been proposed with this purpose. As a rough approximation, these metrics can be classified into three main groups: (i) linear,  (ii) nonlinear and (iii) spectral metrics. While linear metrics, such as the Pearson correlation coefficient, are the most straightforward to be calculated and less time consuming, they suppose the existence of a linear correlation between time series, an assumption that is not fulfilled in the majority of real cases. On the other hand, nonlinear metrics asume a certain nonlinear coupling function $f_n$ between a variable $X$ and a variable $Y$, such as $X=f_n(Y)$. However the estimation of the nonlinear function renders impossible in the majority of cases and certain assumptions have to be assumed for quantifying synchronization. Measures such as the mutual information or the phase locking value are examples of nonlinear metrics, the former assuming a certain statistical interdependency between signals and the latter considering only a phase relation. Finally, spectral metrics, such as the coherence or the imaginary part of coherence, translate the problem to the spectral domain, analyzing the relation between the spectra obtained from the original time series assuming linear/nonlinear relations (see \cite{Pereda2005} for a thorough review about metrics quantifying synchronization in real data sets). 

In the current paper we are concerned about using ordinal patterns, a symbolic representation of temporal data sets, to define a new metric that is able to reveal the synchronization between time series. Bahraminasab \textit{et. al.} \cite{bahraminasab2008} used a symbolic dynamics approach to design a directionality index parameter. Transforming the increment between successive points within a times series into ordinal patterns, authors calculated the mutual information between a process  {$X_1$} at time $t$ and a process {$X_2$} at time $t+\tau$ and next obtained the directionality index as defined in \cite{palus2003}. Applying this methodology to respiratory and cardiac recordings  it is possible to quantify how respiratory oscillations have more influence on cardiac dynamics than vice-versa \cite{bahraminasab2008}. More recently, Li et al. used a similar indicator to evaluate the directionality of the coupling in time series consisting of spikes \cite{li2011}. Using the Izhikevich neuron model \cite{izhikevich2003}, authors showed how that methodology was robust for weak coupling strengths, in the presence of noise or even with multiple pathways of coupling between neurons. More recently, Ros\'ario \textit{et. al.} \cite{rosario2015} used the ordinal patterns observed in EEG datasets, also known as ``motifs" \cite{olofsen2008}, to construct time varying networks and analysed their evolution along time and the properties of the averaged functional network. Specifically, the amount of synchronization between a pair of recorded electrodes of an EEG was obtained by evaluating the number of ordinal patterns co-ocurring at the same time but also at a given lag $\lambda=1$ time steps. Using both positive and negative values of $\lambda$ authors were able to quantify the direction of the interaction between the two time series, i.e., the causality, to further construct temporal time networks. Next, they showed how the resulting time varying functional networks were able to identify those brain regions related to information processing and found differences between healthy individuals and patients suffering from chronic pain \cite{rosario2015}.

In this paper, we also propose the use of symbolic dynamics to evaluate the level of synchronization between time series. However, our methodology consists in a measure of synchronization that does not take into account the existence of a delay time between time series, despite further adaptation to this case is also possible (see Section Conclusions). As in the case of \cite{bahraminasab2008,olofsen2008,rosario2015}, we take advantage of the transformation of a time series into a concatenated series of $D$-dimensional ordinal patterns \cite{Bandt2002} that allow us to quantify the amount of synchronization between two (or more) symbols sequences. The main advantage of our methodology is that it takes into account both the in-phase and anti-phase synchronization of two dynamical systems, the latter being disregarded in the aforementioned proposals based on ordinal patterns. 

We have calculated the $OS$ of two kind of data sets: (i) unidirectionally coupled Lorenz electronic systems and (ii) magnetoencephalographic (MEG) recordings measuring the activity of $241$ sensors placed at the scalp of an individual during resting state. Next, we compared the amount of synchronization computed by \emph{OS} with respect to those obtained from classical metrics like \emph{phase locking value} (PLV), \emph{mutual information} (MI), \emph{spectral coherence} (SC) and \emph{Pearson correlation} (\emph{r}).

\section{Materials and Methods}
\label{S:3}

\subsection{Defining Ordinal Synchronization}

\begin{figure}[h!]
\vspace*{-2cm}
\centering
\includegraphics[width=0.8\linewidth]{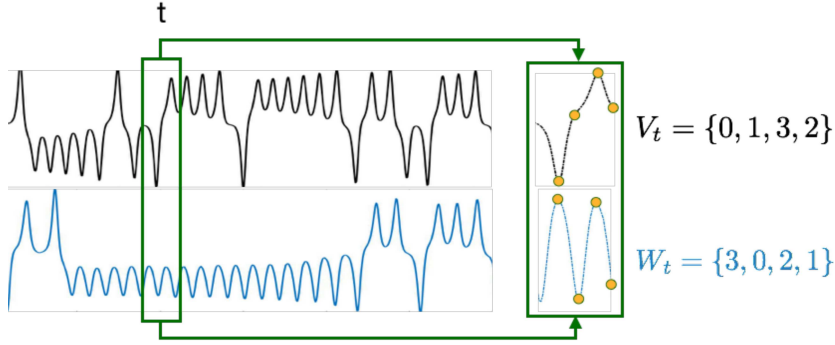}
\caption{Qualitative example of ordinal vectors extraction from two time series. Here $D=4$ is the length of the ordinal patterns. From each time series, an ordinal vector containing the desired number of samples is obtained by ranking its $D$ values at time $t$, inside the vector.}
\label{fig:fig01}
\end{figure}

To compute the (OS) between two time series $X$ and $Y$, we first extract their $D$-dimensional ordinal patterns \cite{Bandt2002}. In this way, we choose a length \textit{D} and divide both time series of length \textit{M} into $L=M/D$ equal segments. Next, we obtain the order of the values included inside each segment, also called the {\it ordinal patterns}:
l
\begin{align}
X_t = \{x_{1},x_{2},\ldots,x_{D}\} &\mapsto V_t = \{v_{1},v_{2},...,v_{D}\}\\
Y_t = \{y_{1},y_{2},\ldots,y_{D}\} &\mapsto W_t = \{w_{1},w_{2},...,w_{D}\}
\end{align}

where $V_t$ and $W_t$ are the {\it ordinal vectors} inside the segment given by $\{t,t+1,..., t+D-1\}$, elements refer to the ordinal position of the values in $X_t$ and $Y_t$, respectively. Note that the elements in $V_t$ and $W_t$ are natural numbers ranging from $0$ to $D-1$. The higher the value in the time series, the higher the corresponding element in the ordinal vector. Following the example depicted in Fig. \ref{fig:fig01}, where $D=4$, we obtain:

\begin{align}
X_t = \{-1.22,0.44,0.91,0.63\} &\mapsto V_t = \{0,1,3,2\}\\
Y_t = \{1.34,0.12,0.78,0.57\} &\mapsto W_t = \{3,0,2,1\}
\end{align}

Then, we take the euclidean norm of each ordinal vector.

\begin{align}
||V_t|| = \sqrt{v_{1}^2+v_{2}^2 + \ldots + v_{D}^2}= \sqrt{0^2+1^2 + \ldots + {(D-1)}^2}
\end{align}

and we call $V^{N}_t=V_t/||V_t||$ and $W^{N}_t=W_t/||W_t||$ the normalized vectors. Note that this step only depends on the length \textit{D}.

Now, we define the raw value of the {\it instantaneous ordinal synchronization} at time $t$ ($IOS^{raw}_t$) as the dot product between both ordinal ordinal vectors $IOS^{raw}_t = \sum_{i=1}^{D}v_{i,t}·w_{i,t}$ (i.e., $V^{N}_t \cdot W^{N}_t$). For a more intuitive interpretation, we linearly rescale the value of $IOS^{raw}_t$ to be bounded between $-1$ and $1$: 

\begin{equation}
IOS_{t} = 2·\left(\frac{IOS^{raw}_{t}-min}{1-min}-0.5\right)
\label{IOSt}
\end{equation}
where $min$ is the minimum possible value of the scalar product between two ordinal vectors. Note that, since the elements of the ordinal vectors are always positive and have only one component equal to zero, the lowest possible scalar product between $V_t$ and $W_t$ is obtained when the order of the elements of vector $V_t$ is inverted in $W_t$. In our example:  

\begin{equation}
min=\frac{0(4)+1(3)+2(2)+3(1)+4(0)}{0^2+1^2+2^2+3^2+4^2}
\end{equation}

In general, for any vector of length \textit{D}:

\begin{equation}
min = \frac{0(D-1)+ 1(D-2) + \ldots + (D-2)1 + (D-1)0}{0^2+1^2 + \ldots + {(D-1)}}
\end{equation}

Following the normalization in \ref{IOSt}), we ensure that two ordinal vectors that follow opposite evolutions will unambiguously lead to a value of $IOS_{t}=-1$, and two vectors whose elements have the same order will have an $IOS_t=1$. 
Being $L = M/D$ the total number of ordinal vectors in time series of $M$ points, the final value of the ordinal synchronization $OS\{X,Y\}$ for a given pair of time series $X$ and $Y$ is obtained averaging the instantaneous values of $IOS_t$ along the whole time series:

\begin{equation}
OS\{X,Y\}= \langle IOS_t \rangle
\label{osdefx}
\end{equation}

Since we consider the $IOS_t$ of consecutive (i.e., non-overlapping) time windows, the value of $t$ in Eq. \ref{osdefx} is given by the expression $t=1+i D$, with $i$ being a natural number bounded by $0 \leq i \leq L-1$. Note that it is also possible to define a {\it sliding} $OS$ just by increasing $t$ in one unit for every $IOS_t$ instead of considering consecutive windows.

\subsection{Experimental results: Electronic Lorenz Systems}
 
\begin{figure}[h!]
	\centering
	\includegraphics[scale=0.4]{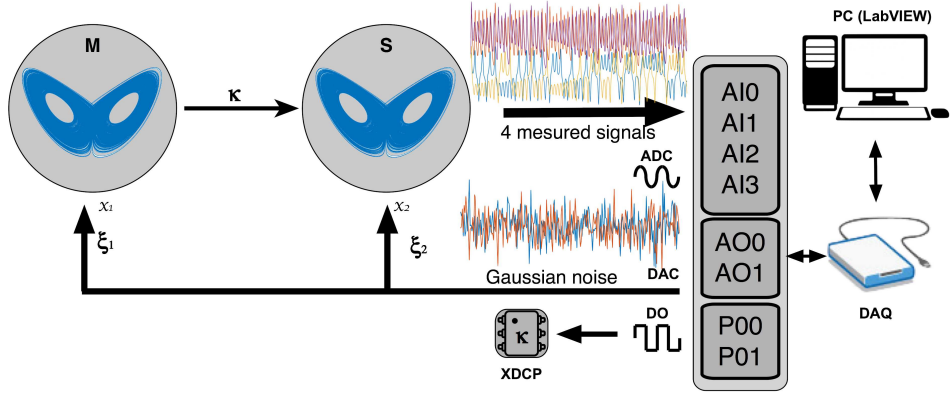}
		\caption{Schematic representation of Lorenz systems in a master-slave configuration. Signals are measured through a DAQ card (Ports AI0-AI3) and stored in a PC. Digital output (P00-P01) ports and XDCP control the value of coupling strength $\kappa$. Analog output ports (AO0-AO1) introduce the external noise signals $\xi_1$ and $\xi_2$ perturbing the $x_1$ master (M) and $x_2$ slave (S) variables, respectively.}
	\label{fig:fig02}
\end{figure}

We analyzed the transition to the synchronized regime of two coupled Lorenz oscillators \cite{lorenz1963}. We implemented an electronic version of the Lorenz system, whose equations are detailed in Appendix B. Two Lorenz circuits are coupled unidirectionally in a master-slave configuration (see Fig. \ref{fig:fig02}) with a coupling strength $\kappa$ that can be modified. Our experiments include two conditions: in the first one, $\kappa$ is modified in the absence of external noise; in the second one, $\kappa$ varies in presence of Gaussian noise with band selection. 
The (AI0-AI3) input ports of a data acquisition (DAQ) card are used for sampling the $x$ and $z$ variables of each circuit, while the output ports AO0 and AO1 generate two 
different noise signals ($\xi_1$, $\xi_2$) that perturb the dynamics of the Lorenz circuits through variable $x$ of each circuit. In this way, an external source of noise can be introduced to check the robustness of the experiments.
The circuit responsible of the coupling strength $\kappa$ is controlled by a digital potentiometer XDCP, which is adjusted by digital pulses from ports P00 and P01. Noisy signals were designed in LabVIEW, using a Gaussian White Noise library \cite{Ni_noise} that generates two different Gaussian-distributed pseudorandom sequences bounded 
 between [-1 1]. All the experimental process is controlled by a virtual interface in LabVIEW 2016 (PC).

The experiment works in the following way: First, $\kappa$ is set to zero and digital pulses (P00 and P01) are sent to the digital potentiometer until the highest value of $\kappa$ is reached. Second, variables $x$ and $z$ of the circuits are acquired by the analog ports (AI0-AI3) in order to compute the synchronization metrics. Initially, we have obtained all results for $\xi_1=\xi_2 = 0$, i.e., in the absence of external noise, and then, after a moderate amount of noise is introduced, all synchronization metrics are calculated again (See Appendix C). Every signal, with or without noise, has a length of 30000 samples.

\subsection{Applications to magnetoencephalographic recordings}

We have checked the performance of the $OS$ in the context of neuroscientific datasets. Specifically, we quantified the level of synchronization between pairs of channels of MEG recordings. Data sets have been obtained from the Human Connectome Project (for details, see \cite{larson2013} and https://www.humanconnectome.org). The experimental data sets consist of $30$ MEG recordings of an individual during resting state for a period of approximately $2$ minutes each. During the scan, the subject were supine and maintained fixation on a projected red crosshair on a dark background. Brain activity was scanned with 241 magnetometers on a whole head MAGNES $3600$ (4D Neuroimaging, San Diego, CA, USA) system housed in a magnetically shielded room. The root-mean-squared noise of the magnetometers is about $5$ fT/sqrt ($Hz$) on average in the white-noise range (above 2 $Hz$). Data was recorded at sampling rate of $f_s \approx 508.63$ $Hz$. Five current coils attached to the subject, in combination with structural-imaging data and head-surface tracings, were used to localize the brain in geometric relation to the magnetometers and to monitor and partially correct for head movement during the MEG acquisition. Artifacts, bad channels, and bad segments were identified and removed from the MEG recordings, which were processed with a pipeline based on independent component analysis to identify and clean environmental and subject's artifacts \cite{larson2013}.

\section{Results}  
  
\subsection{Nonlinear electronic circuits}

In order to assess the validity of $OS$, we have explored its performance for different values of $D$, from 3 to the full length of the time series under evaluation. Since it is the first time $OS$ is used, we have compared it to classical measures of correlation, namely Pearson correlation coefficient ($r$), spectral coherence ($SC$), phase locking value ($PLV$) and mutual information ($MI$). We have used two kinds of data sets to validate $OS$, on the one hand, experimental time series from nonlinear electronic circuits, and on the other hand, MEG recordings.

\begin{figure}[h!]
	\begin{center}
		\includegraphics[width=0.45\linewidth]{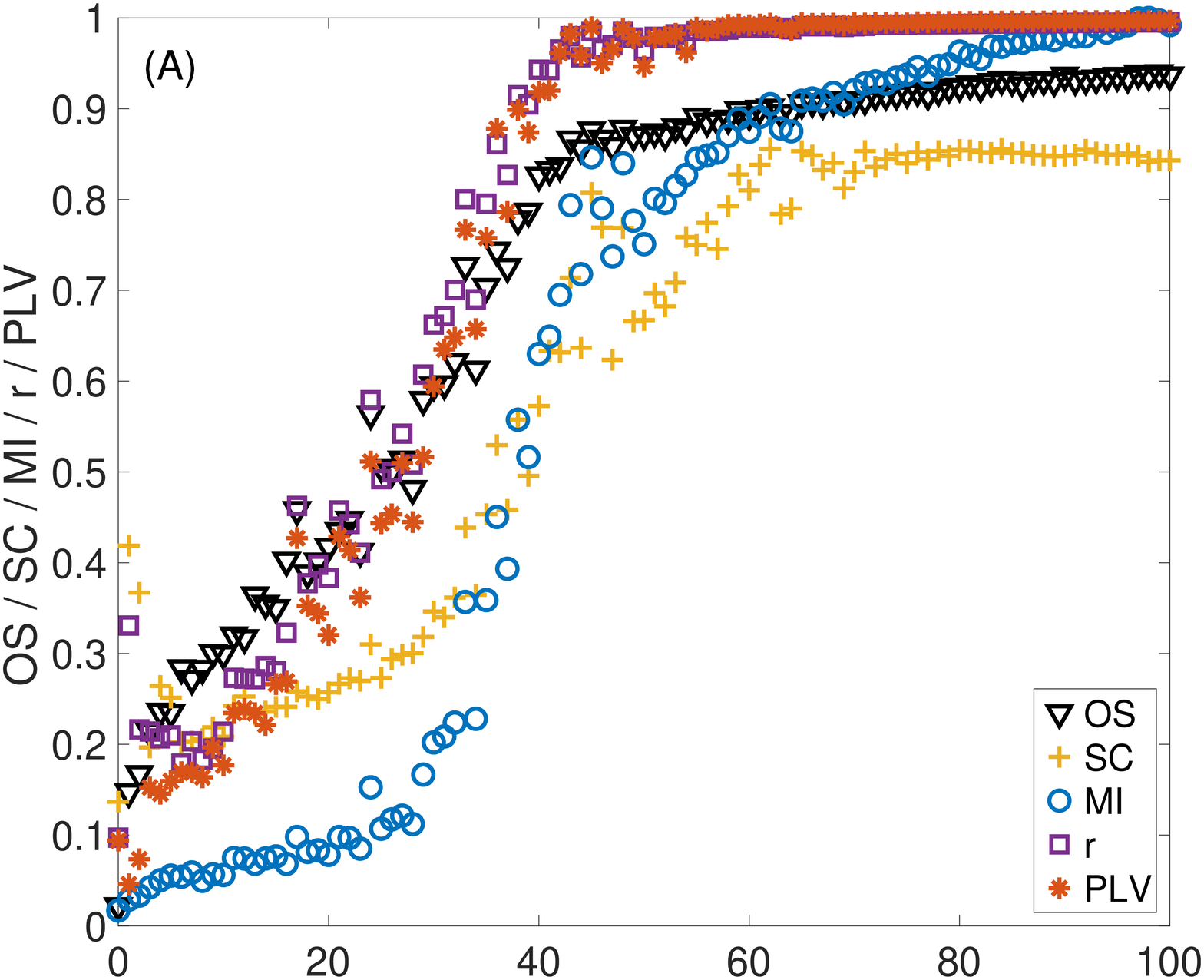}\hspace{0.5cm}
		\includegraphics[width=0.45\linewidth]{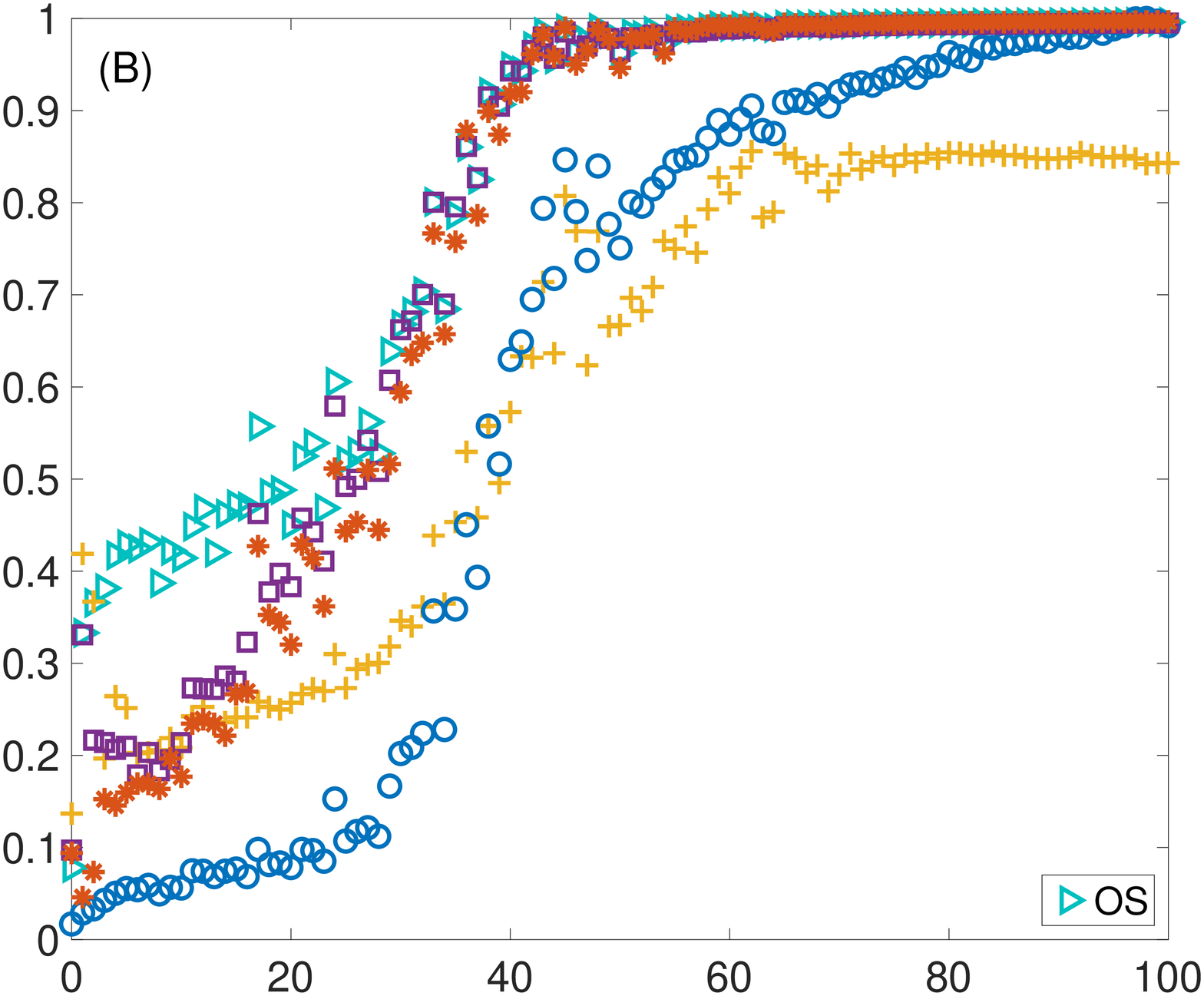}\vspace{0.35cm}
		\includegraphics[width=0.45\linewidth]{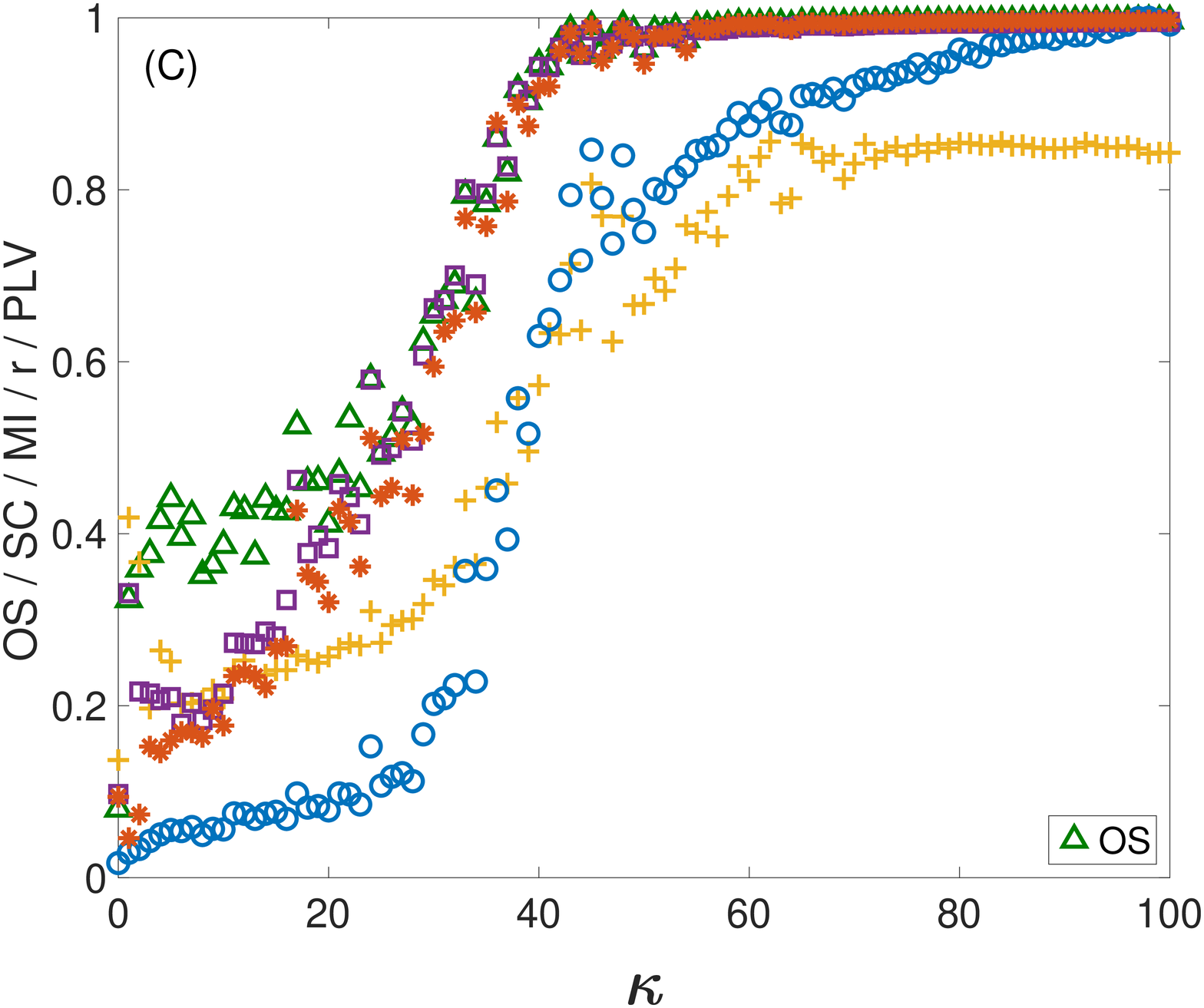}\hspace{0.5cm}
		\includegraphics[width=0.45\linewidth]{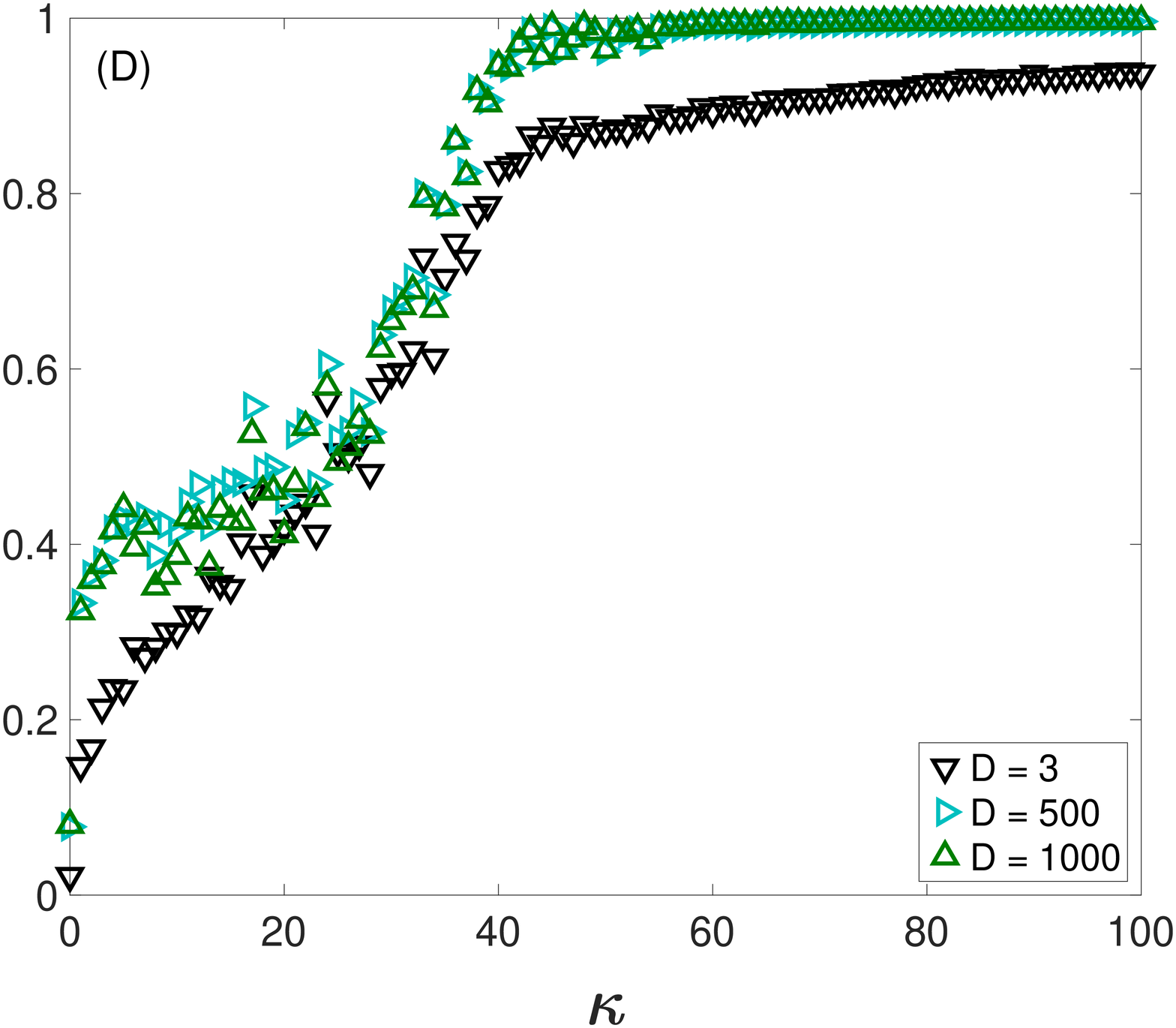}				
		\caption{Synchronization against coupling strength $\kappa$ as measured with $PLV$ (red stars), $SC$ (yellow crosses), $MI$ (light blue circles), $r$ (purple squares). $OS$ (triangles) is plotted for $D = 3$ (A) (black downward-pointing), $D=500$ (B) (turquoise right-pointing) and $D=1000$ (C) (green upward-pointing). For comparison purposes, plot (D) shows $OS$  against $\kappa$ for different vector lengths, $D=3$, $D=500$ and $D=1000$.}
\label{fig:fig03}
	\end{center}	
\end{figure}

First, we take advantage of the ability of controlling the coupling strength between electronic circuits and investigate how $OS$ changes as two dynamical systems smoothly vary their level of synchronization from being unsynchronized to completely synchronized.
Specifically, two electronic Lorenz systems are unidirectionally coupled with a parameter $\kappa$ controlling their coupling strength (see Appendix B for details). Initially, we do not perturb the oscillators with external noise (see Appendix C for the case of including external noisy signals). However, we can not avoid the intrinsic noise of the electronic circuits together with the tolerance of the electronic components (between 5 \% and 10 \%). Figure \ref{fig:fig03} shows how the value of $OS$ changes as the coupling strength $\kappa$ is increased from zero. Since the value of $OS$ depends on the length of the ordinal vectors, we show the results for three different values: $D = 3$ (Fig.  \ref{fig:fig03}A), $D = 500$ (Fig.  \ref{fig:fig03}B) and $D = 1000$ (Fig.  \ref{fig:fig03}C). Note that, by increasing the length of the vectors, we are obtaining the amount of synchronization at different time scales. Together with $OS$, we plot the values of the rest of synchronization metrics in (A), (B) and (C), which remain unaltered in the three plots (since they do not depend on $D$).

In all cases, we observe that $OS$ increases for low to moderate values of $\kappa$ and remains at a high value once a certain threshold is reached. This behaviour is similar to the rest of the synchronization metrics. However, both $MI$ and $SC$ seem to saturate at values of $\kappa$ higher than $r$, $PLV$ and $OS$, which seem to reach a plateau around $\kappa=40$. Figure  \ref{fig:fig03}D shows the comparison of $OS$ for the three different values of $D$. Here, we can also observe how at $D=3$, $OS$ has a different qualitative behaviour from $D=500$ and $D=1000$, since it stays around 0.9 and does not reach 1 as in the windows of longer lengths. The reason is the existence of intrinsic noise of the electronic circuits, that affects much more the alignment of the ordinal vectors of shorter lengths than those with higher dimensions.

\begin{figure}[h!]
	\begin{center}
		\includegraphics[width=0.4\linewidth]{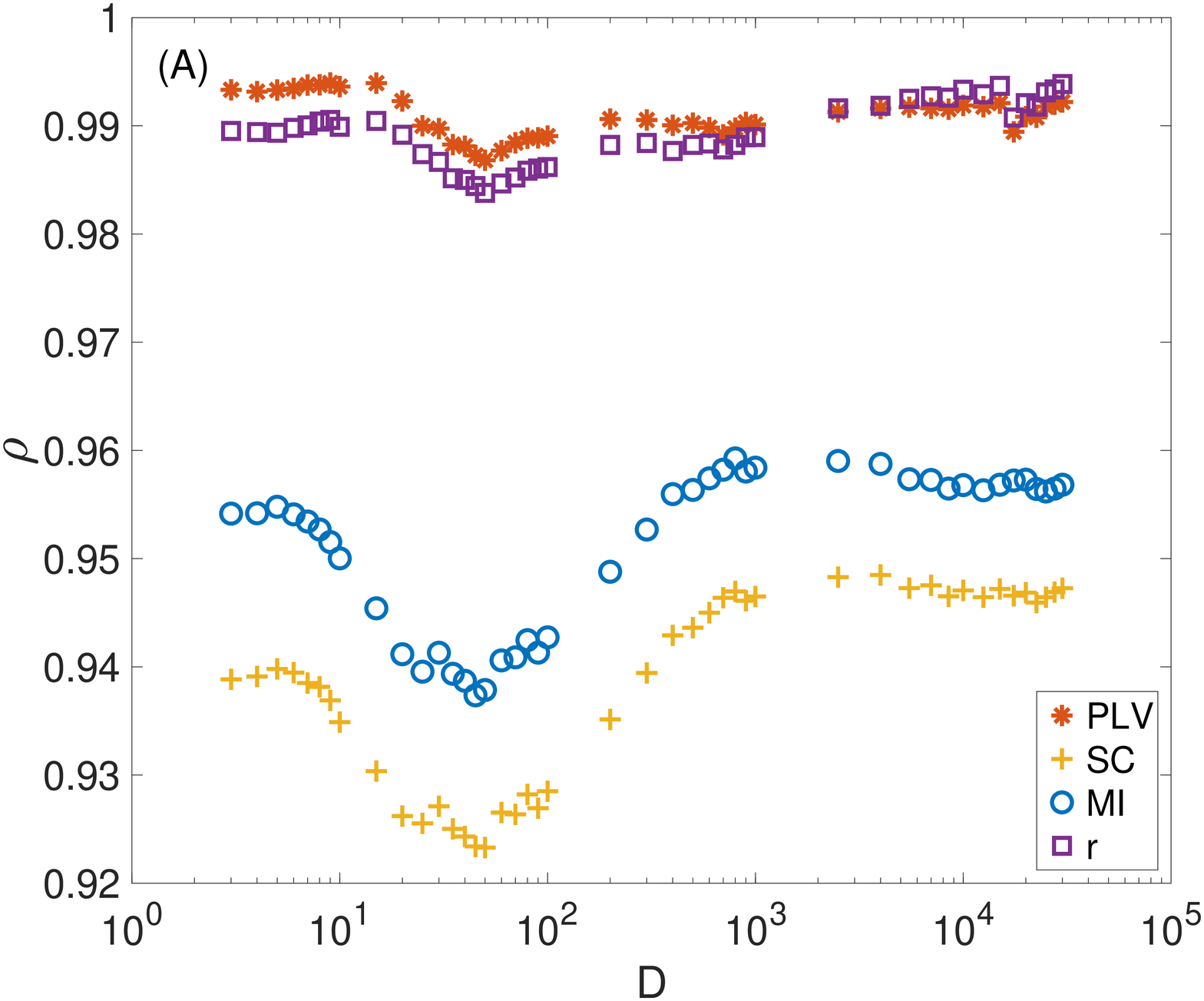}\hspace{0.5cm}
		\includegraphics[width=0.4\linewidth]{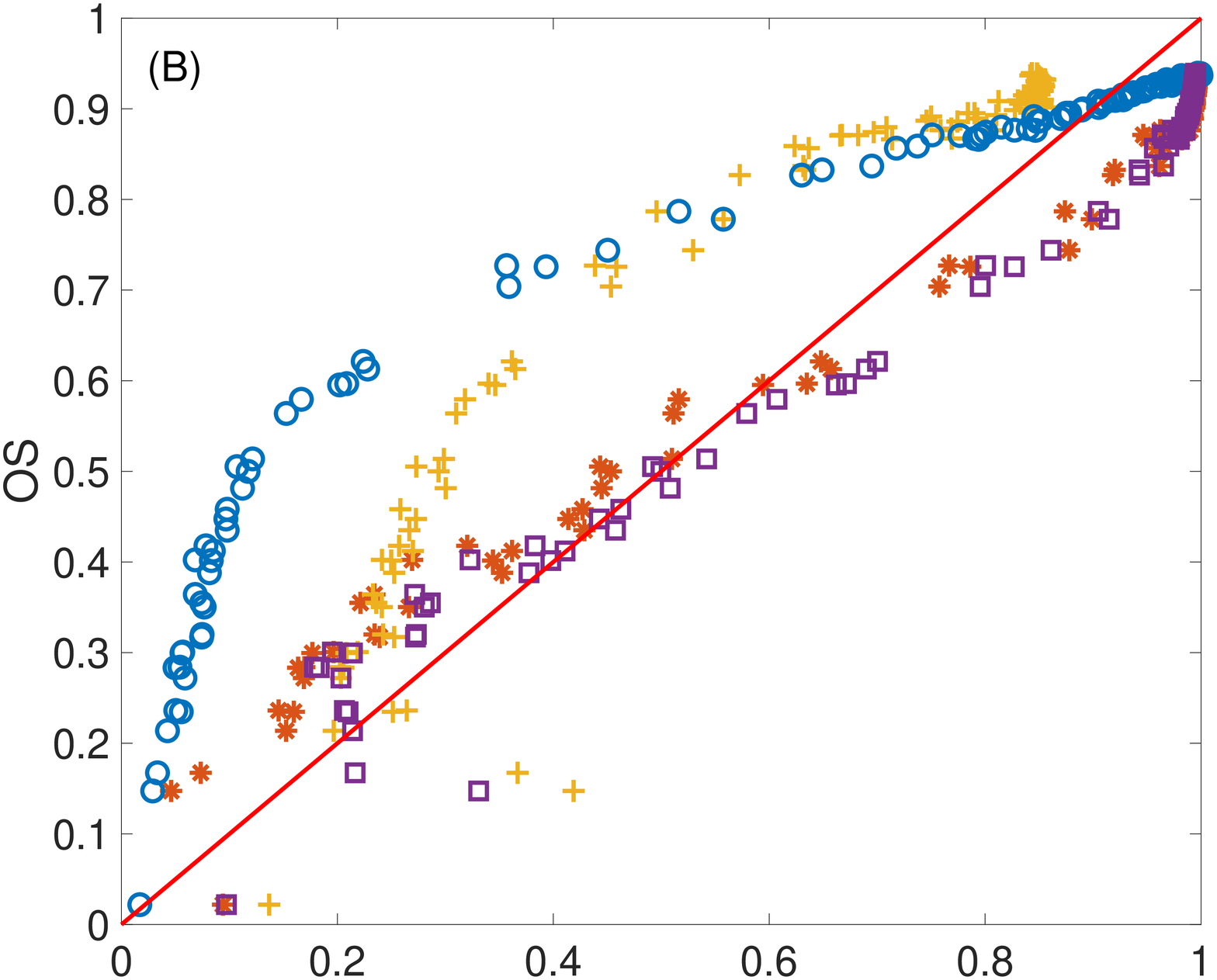}\vspace{0.35cm}
		\includegraphics[width=0.4\linewidth]{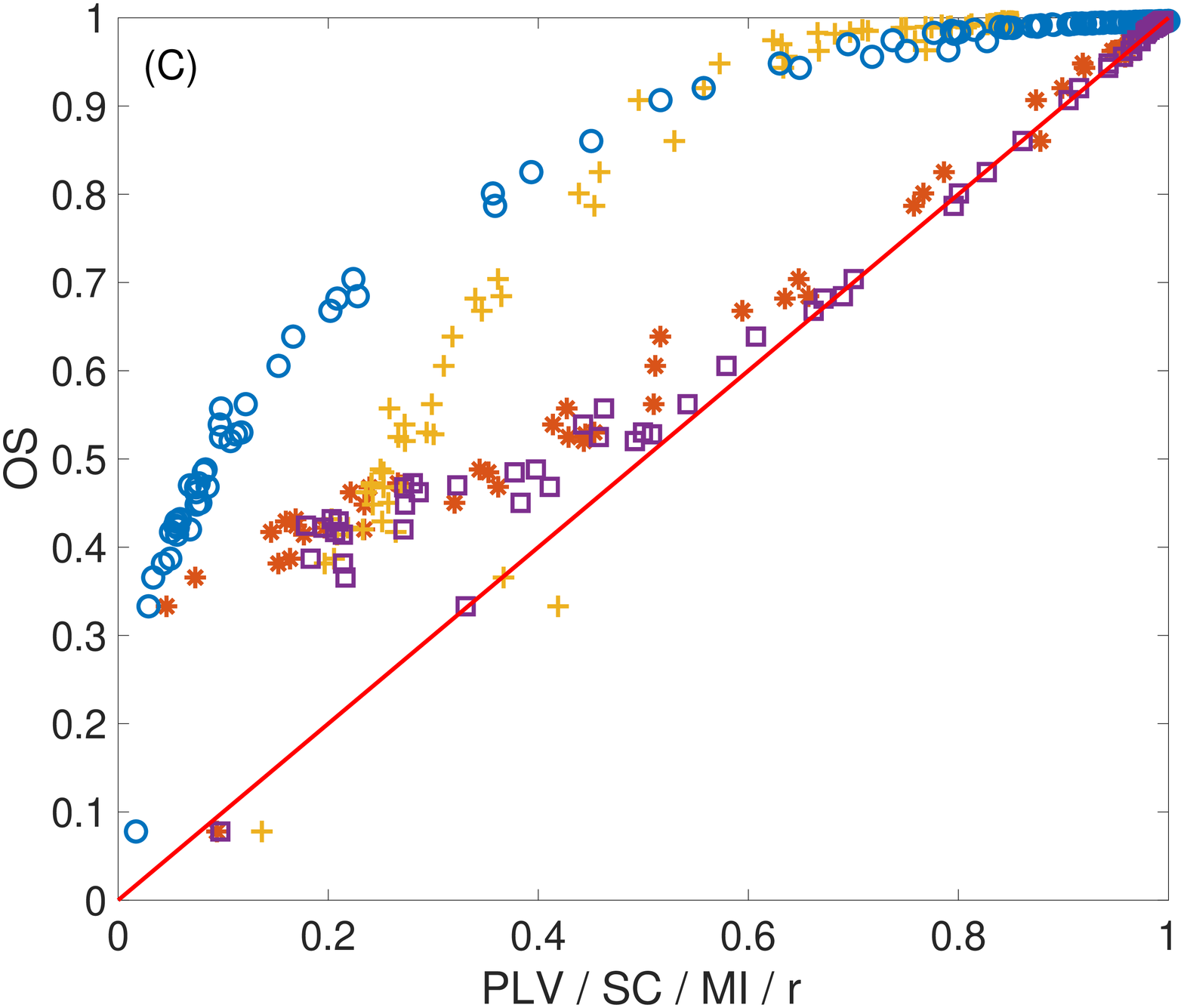}\hspace{0.5cm}	
		\includegraphics[width=0.4\linewidth]{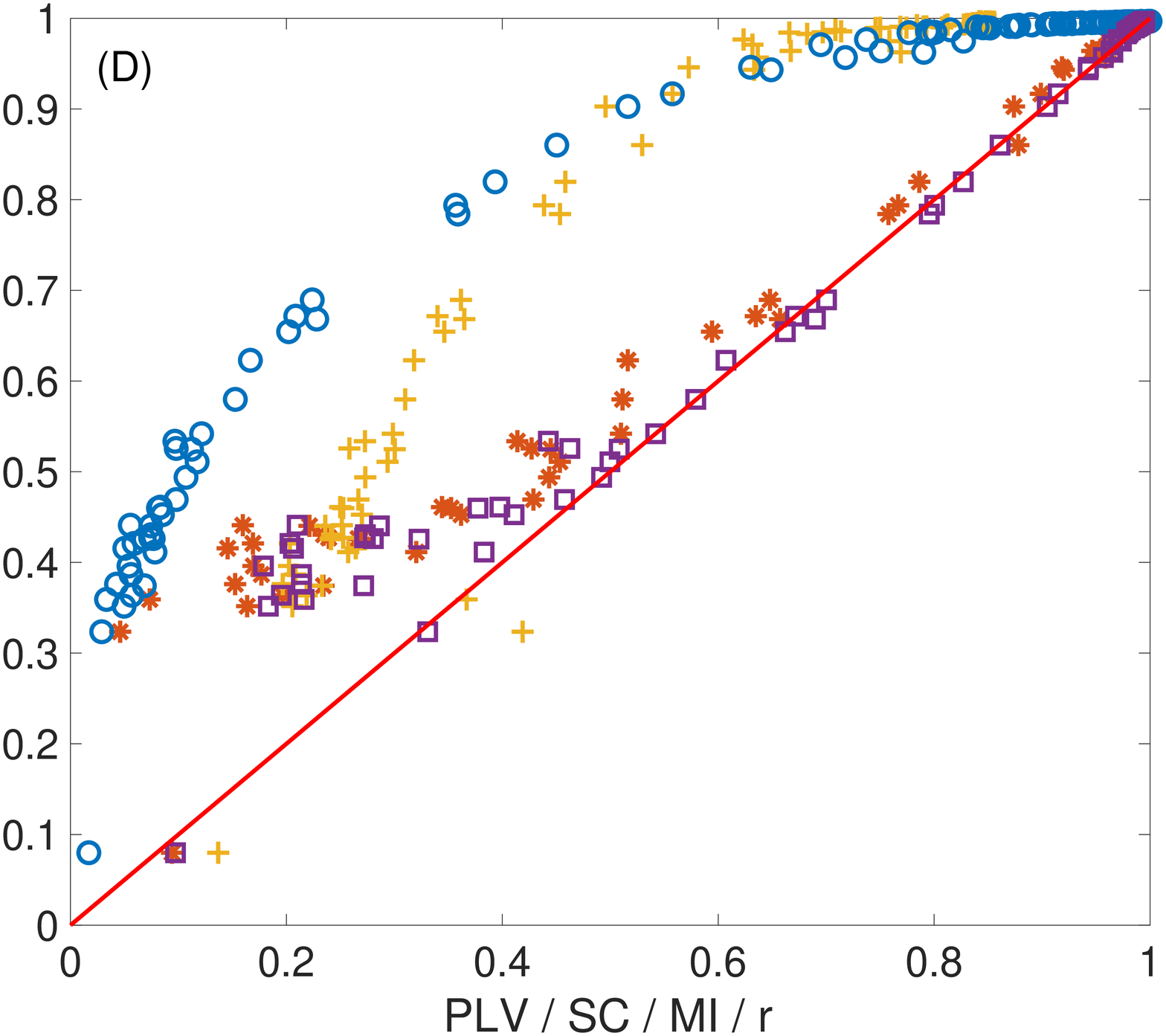}	
    \caption{Panel (A) shows the average correlation ($\rho$) between each synchronization measure and OS depending on the vector length $D$. Panels (B)-(D) show the correlation between $OS$ and all other synchronization measures varying the coupling strength (from 0 to 100), for $D = 3$ (B), $D  = 500$ (C) and $D = 1000$ (D).Following the same notation as in Fig. \ref{fig:fig03}, synchronization measures are (MI; blue), (r; purple), (SC; yellow) and (PLV; red). The red line corresponds to $y=x$. }
\label{fig:fig04}								
	\end{center}
\end{figure}   

Figure \ref{fig:fig04} shows the average correlation ($\rho$) between each D-dependent $OS$ and the rest of synchronization metrics with zero noise. Note that correlations are higher than $0.92$ in all cases, although it seems to be certain vector lengths that maximize these correlations. Also note that correlations with $PLV$ and $r$ are the highest and, in all cases, very close to 1. At the same time, $MI$ and $SC$ show lower correlations that, in turn, seem to be more dependent on the value of the vector length $D$ (see Fig. \ref{fig:fig04}A).

We can investigate how $OS$ is related to the rest of the synchronization metrics in more detail by setting the length of the ordinal vectors to a given value ($3$, $500$ or $1000$ in this case) and observe the influence of the level of synchronization (Fig. \ref{fig:fig04}B, C and D). For any of the three selected lengths, $OS$ shows a linear relation with $PLV$ and $r$, especially at values of $OS$ higher than $0.5$. However, the relation with $SC$ and $MI$ seems to be nonlinear in all cases. Interestingly, for low levels of synchronization, $OS$ increases much faster than these two latter metrics. While $SC$ saturates around $0.8$, $MI$ finally increases faster than $OS$ only for high values of synchronization, eventually reaching the value of $OS$ around 1. Also note how, in the case of $D=3$  (Fig. \ref{fig:fig04}B), the intrinsic noise of the electronic circuits prevents $OS$ to reach the value of one. This behaviour can be observed even clearer in the case of adding more noise into the system, as shown in Appendix C.

\subsection{MEG signals}

\begin{figure}[h!]
\centering
\includegraphics[width=0.35\linewidth]{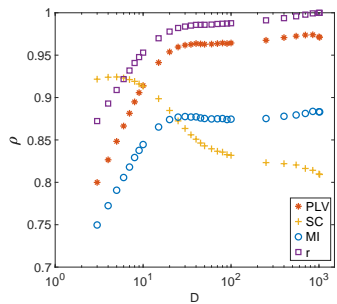}
\caption{Correlation $\rho$ between different synchronization metrics and $OS$ as a function of the length $D$ of the ordinal vectors. As in previous figures: PLV (red), SC (yellow), MI (blue), $r$ (purple).}
\label{fig:fig05}
\end{figure}

The second application is the evaluation of the level of synchronization between the 241 sensors measuring the activity of an individual during resting state. Concretely, we have 30 recordings of 2 minutes each. In this case, we can not control the amount of coupling between sensors but, alternatively, we have a diversity of levels of synchronizations between all possible pairs of sensors. Figure \ref{fig:fig05} shows how the correlations between $OS$ and the rest of the metrics change depending on $D$. As we can observe, correlations are high in all cases except for \emph{SC}, but this one saturates around the same $D$ as the other synchronizations does.

 As in the case of the electronic Lorenz oscillators, tuning the value of $D$ allows to obtain values of $OS$ closer, or more correlated, to other metrics. In fact, two different regions are clearly observed: (i) for values of $D\leq20$ the correlation of $OS$ with PLV, r, and MI increases with $D$, while (ii) for $D>20$ correlation saturates around the highest value, being $r$ the metric with the highest correlation. Interestingly, the behaviour of the SC goes in the opposite direction, decreasing for higher values of $D$.

\begin{figure}[h!]
	\begin{center}
			\includegraphics[width=0.3\linewidth,keepaspectratio]{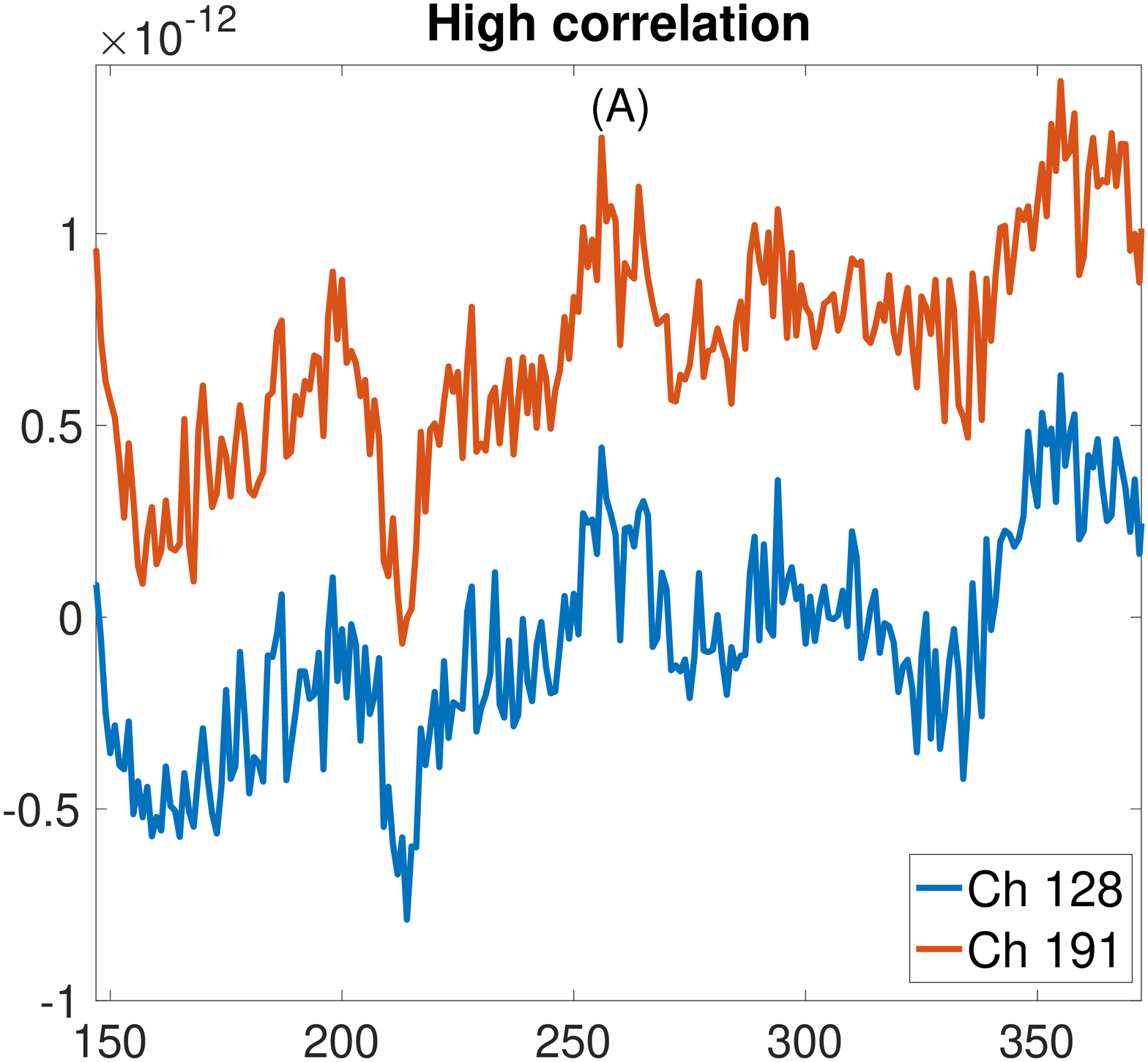}
			\includegraphics[width=0.3\linewidth,keepaspectratio]{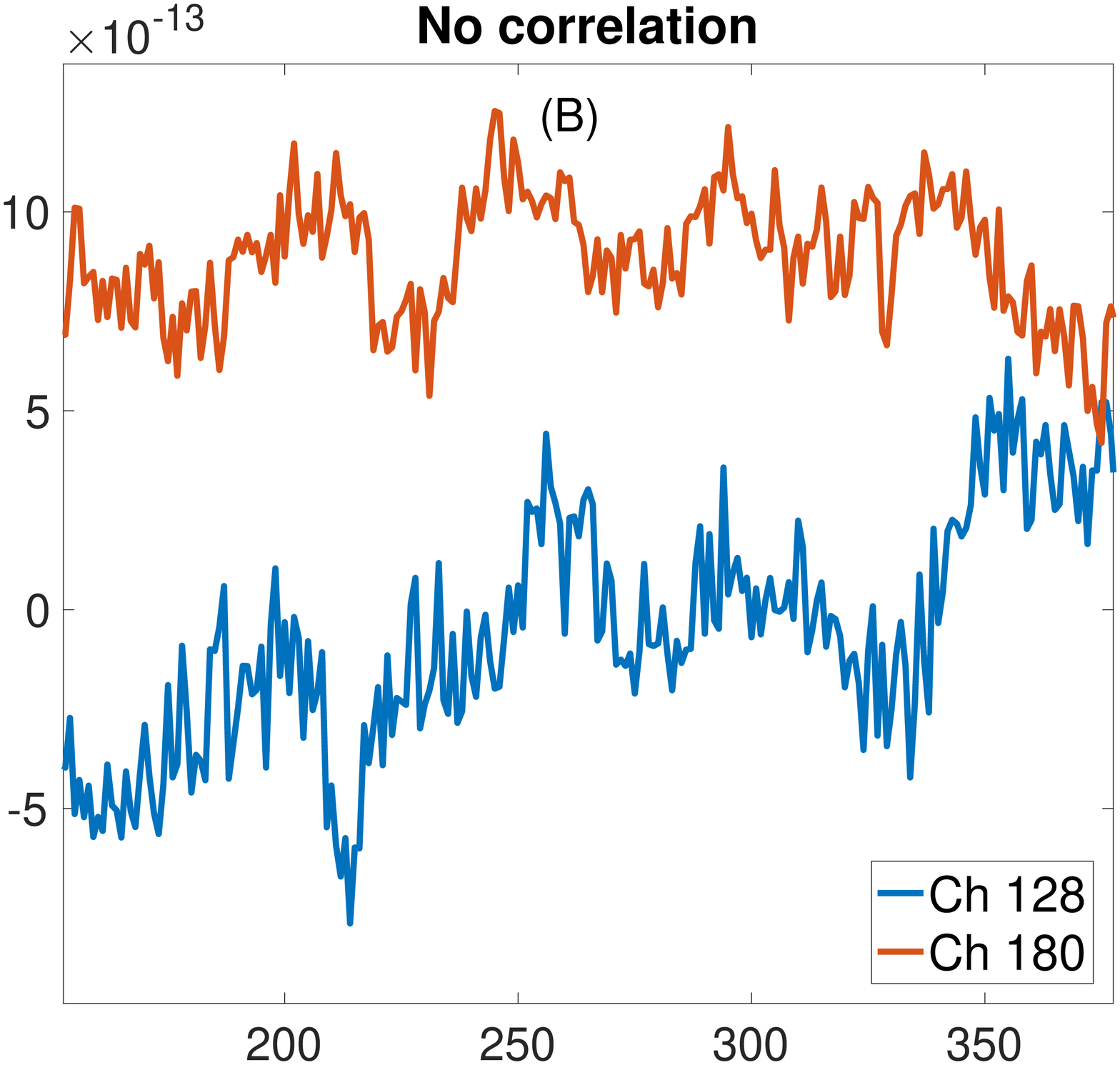}
			\includegraphics[width=0.3\linewidth,keepaspectratio]{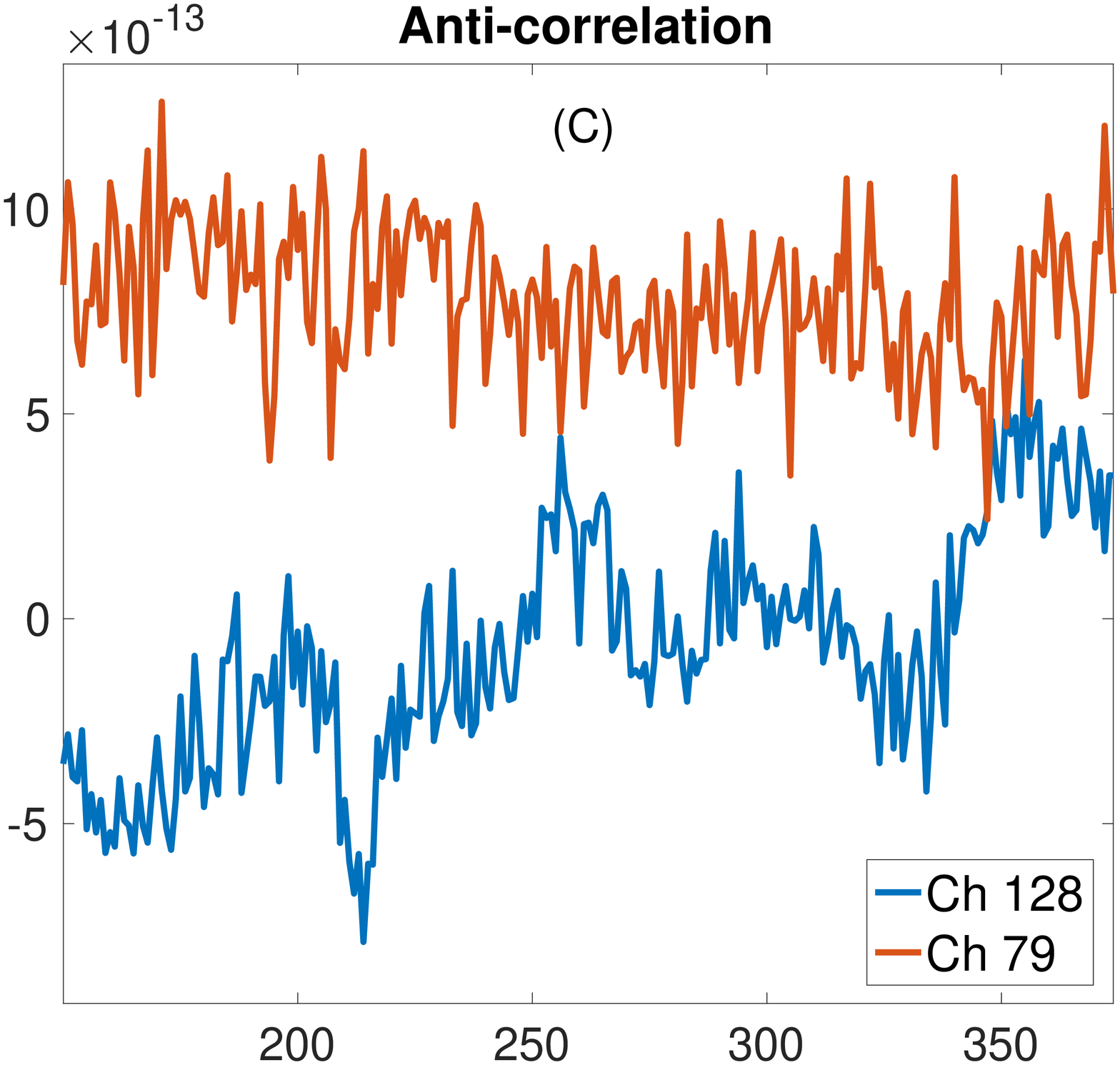}\vspace{0.35cm}
			\includegraphics[width=0.3\linewidth,keepaspectratio]{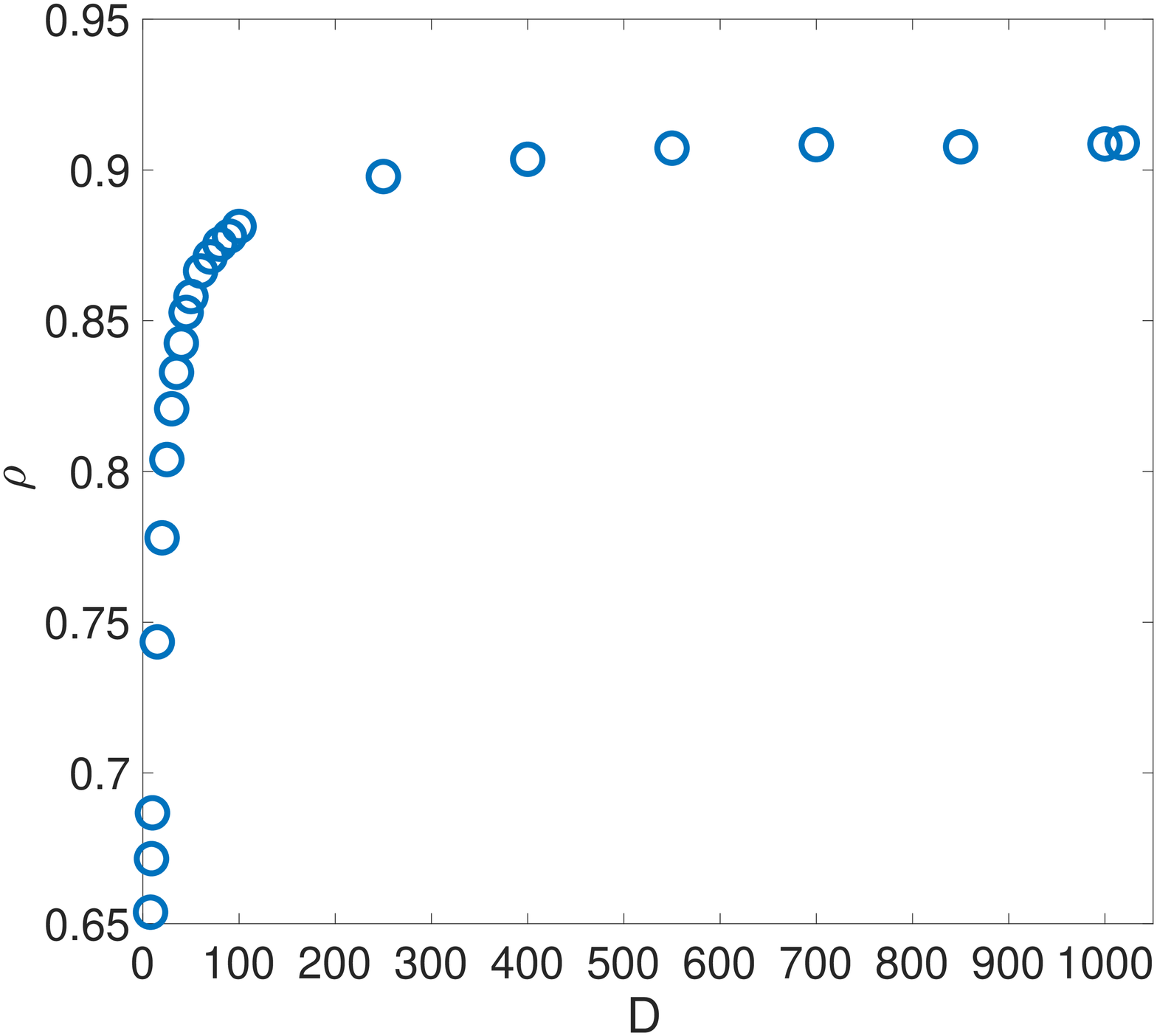}
			\includegraphics[width=0.3\linewidth,keepaspectratio]{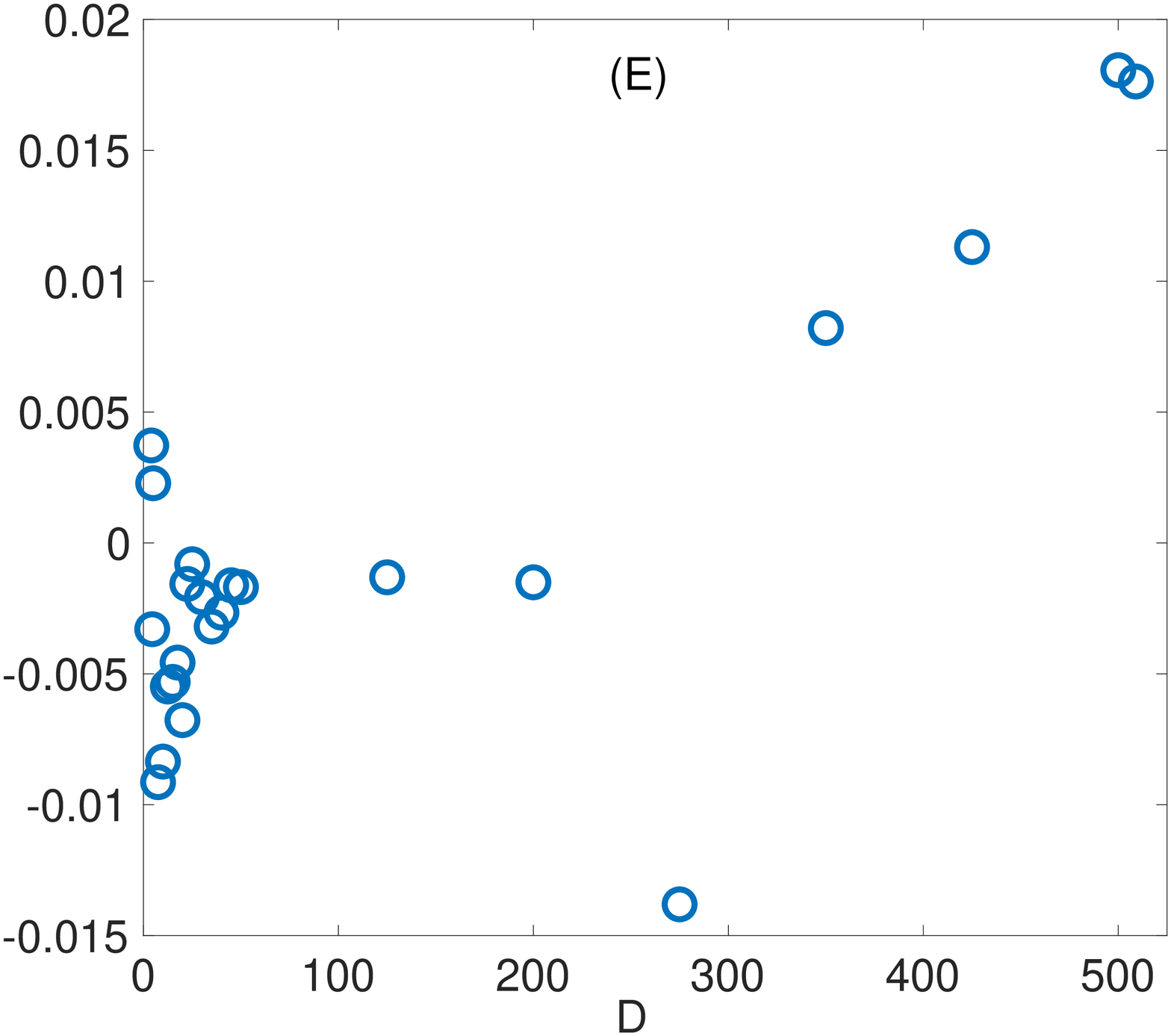}
			\includegraphics[width=0.3\linewidth,keepaspectratio]{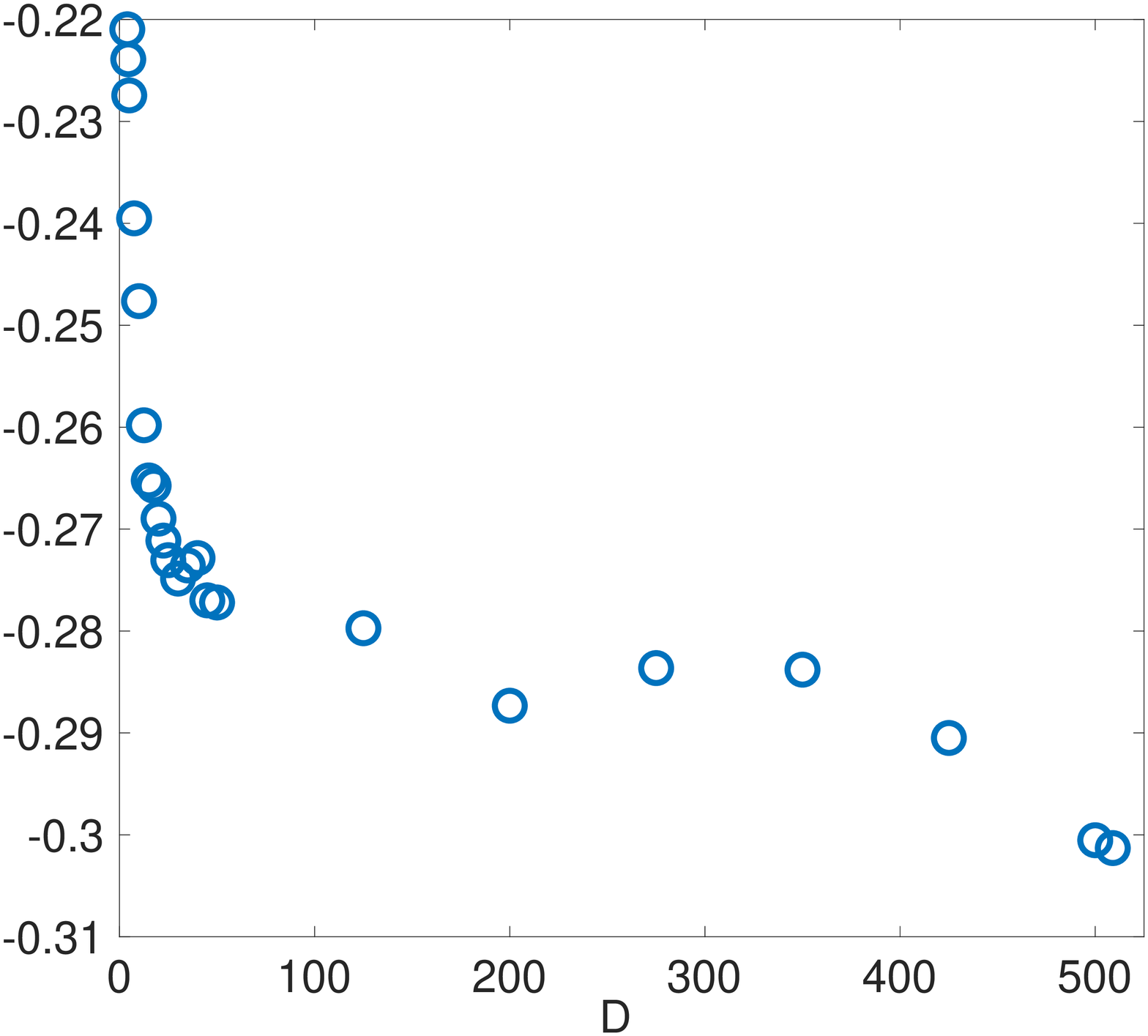}			    
			\caption{Example of the MEG time series and their corresponding $OS$ vs $D$ for three different situations: two sensors with high correlation (A and D), no correlation (B and E) and anti-correlated (C and F).  Upper panel shows part of the raw signals recorded at the sensors while bottom panel shows $OS$ depending on the length $D$.}
\label{fig:fig06}
	\end{center}
\end{figure}

In order to gain insights about how the behaviour of $OS$ depends on the level of synchronization and the length $D$, we plot 
three different cases in Figure \ref{fig:fig06}. In (A), we show the time series of two highly-correlated sensors, with their 
corresponding $OS$ value depending on $D$ (Fig. \ref{fig:fig06}D). Plot (B) and (C) show the cases of two 
uncorrelated and negatively correlated sensors, respectively, with their values of $OS$ (Fig. \ref{fig:fig06}E and F). 
Note that for the positive (negative) case, correlations tend to stabilize as $D$ grows, indicating the existence 
of a certain temporal scale at which synchronization is increased (reduced). Also note that, when time series are 
not correlated, this pattern is not that clear, and $OS$ values remain low for any value of $D$.

\begin{figure}[h!]
\vspace*{-1cm}
\hspace*{-0.2cm}
	\begin{subfigure}[h!]{.8\linewidth}
\includegraphics[width=1.2\linewidth, height=0.3\linewidth]{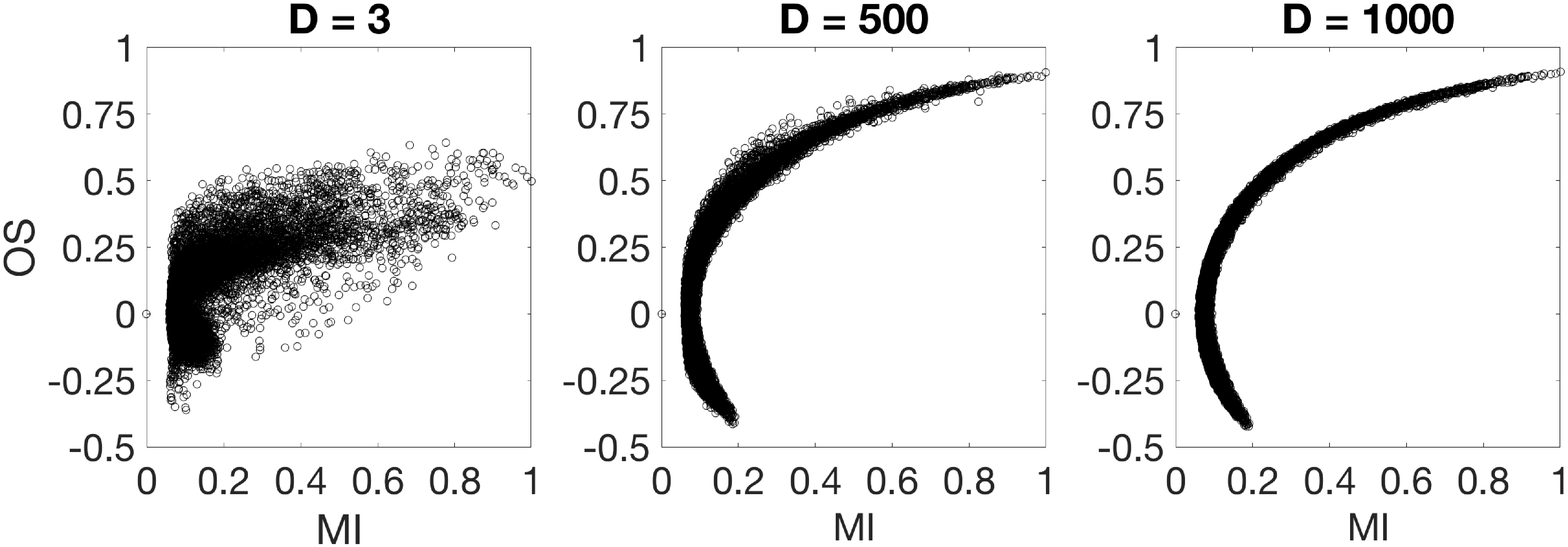}
	\end{subfigure}\\
\hspace*{-0.2cm}
	\begin{subfigure}[h!]{.8\linewidth}
\includegraphics[width=1.2\linewidth, height=0.3\linewidth]{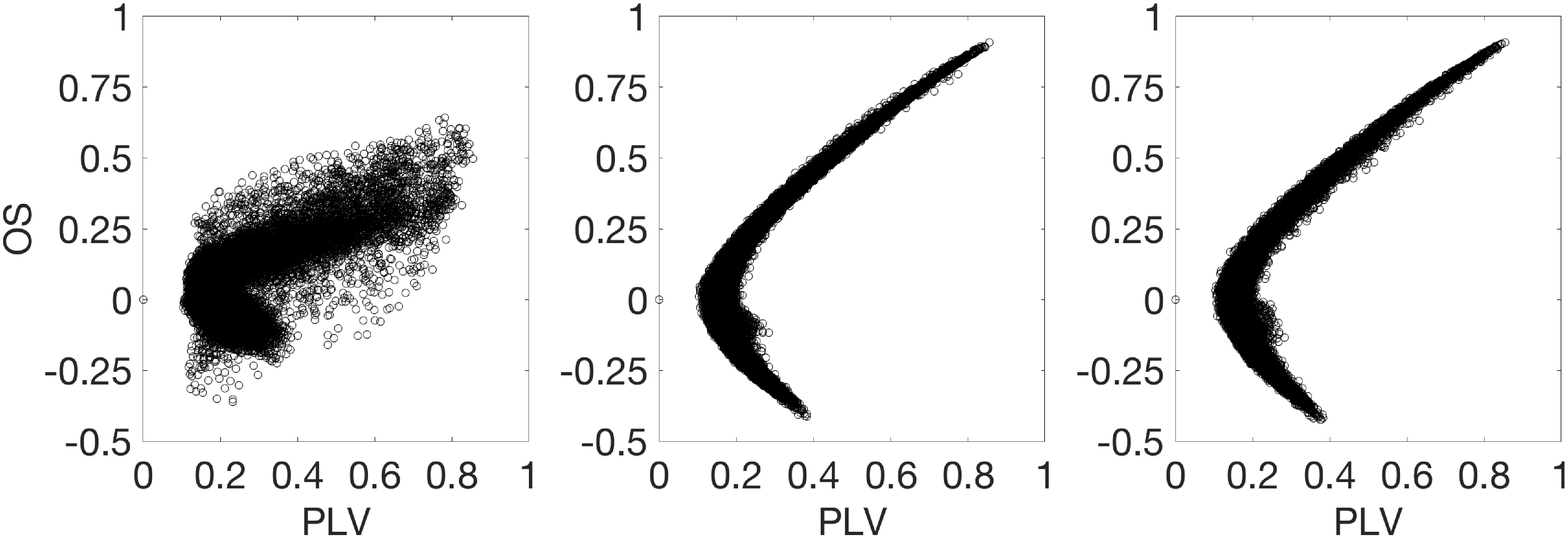}
	\end{subfigure}\\
\hspace*{-0.2cm}
    \begin{subfigure}[h!]{.8\linewidth}
\includegraphics[width=1.2\linewidth, height=0.3\linewidth]{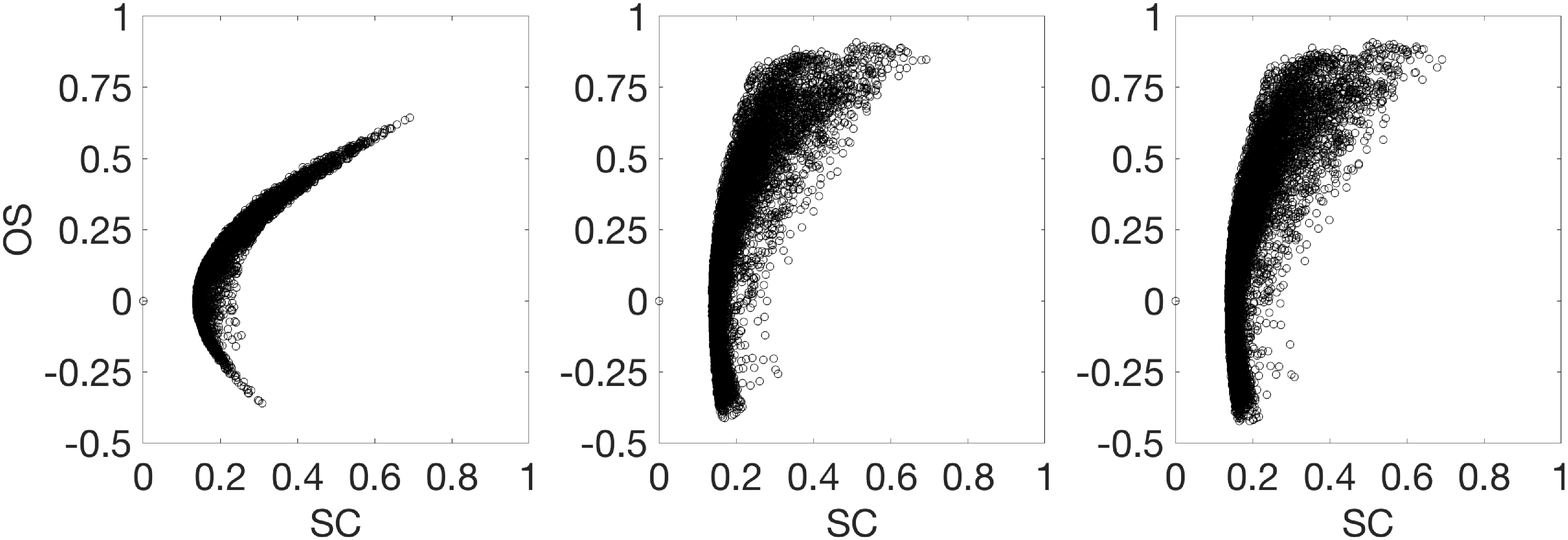}
	\end{subfigure}\\
\hspace*{-0.2cm}
	\begin{subfigure}[h!]{.8\linewidth}
\includegraphics[width=1.2\linewidth, height=0.3\linewidth]{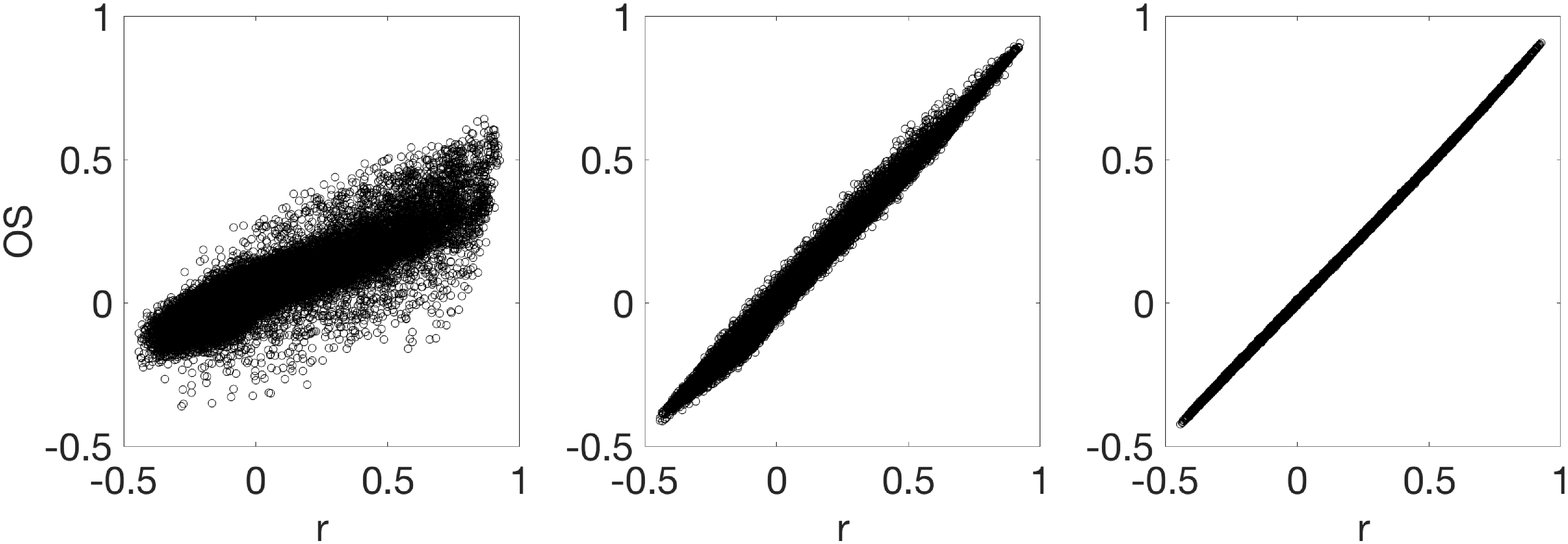}
	\end{subfigure}
    \caption{Correlation between $OS$ and other synchronization metrics in MEG data sets: $MI$ (upper row), $PLV$ (second row), $SC$ (third row) and $r$ (bottom row). Each column corresponds to an $OS$ obtained with different lengths $D$ of the ordinal vectors: $D=3$ (left column), $D=500$ (middle column) and $D=1000$ (right column).} 
\label{fig:fig07}
\end{figure}

We had analyzed the relation of $OS$ with the rest of the metrics according to the level of synchronization. Figure \ref{fig:fig07} shows a panel of plots capturing the correlations between $OS$ and all other synchronization metrics for the MEG signals. Left plots show the case of $D = 3$, middle plots show $D = 500$ and right plots show $D = 1000$. Different conclusions can be drawn depending on the synchronization metric $OS$ is compared to. In the case of $MI$ (first row), the existence of a nonlinear correlation between both metrics arises. However, this correlation decreases with the length of the ordinal vectors, becoming rather noisy for $D=3$. This behaviour is induced by the intrinsic noise of the MEG signals that, as in the case of electronic circuits, affects the value of $OS$ when short lengths of the ordinal vectors are considered. Also note that $MI$ is not able to distinguish positive from negative correlations between time series, a fact that makes $OS$ an interesting metric when both kind of synchronizations are expected. In our case, for example, despite the highest values of $OS$ are close to $1$, the lowest ones arrive to $-0.35$, indicating the existence of anti-correlated dynamics between certain pairs of sensors. A similar behaviour is reported in the case of the comparison with $PLV$ (second row). Again, a nonlinear relation exists between both metrics, which is rather noisy at low values of the ordinal vector lengths ($D=3$). $PLV$ has also the same limitations as $MI$, since it does not differentiate between positive and negative correlations. Interestingly, the relation with $SC$ is 
different from the two previous metrics (third row). Despite a nonlinear correlation between $OS$ and $SC$ seems to be present in the plots, this correlation is deteriorated with the increase of $D$.

Finally, $OS$ shows a clear linear correlation with $r$ (bottom row), which, as in the case of $MI$ and $PLV$ becomes noisy for low values of $D$. Note that, the loss of correlation for low values of $D$ is indicating that, at short time scales, $OS$ is capturing a different pattern of synchronization than at large scales. This is an interesting feature of $OS$ which suggests that, when using it as a metric to evaluate synchronization between signals, it is appealing to carry out an analysis depending on the vector length in order to reveal the existence of different levels of synchronization at different time scales.

\section{Conclusions}
\label{S.5}

We have introduced the {\it Ordinal Synchronization} ($OS$), a new metric to evaluate the level of synchronization between time series by means of a projection into ordinal patterns. We have checked the performance of OS with two kinds of experimental data sets obtained from: (i) unidirectionally coupled nonlinear electronic circuits and (ii) 30 magnetoencephalographic recordings containing the signals of 241 channels. There are several advantages of using $OS$. First, it is able to capture in-phase and anti-phase synchronization. Second, tuning the length of the ordinal vectors $D$, it is possible to evaluate the level of synchronization at different time scales. Third, it is not necessary to assume any {\it a priori} property of the time series, such as stationarity or linear coupling. Fourth, the calculation of $OS$ is extremely fast, especially when compared with other metrics such as $MI$. On the other hand, we have also seen that one of the elements affecting the value of $OS$ is the existence of noise, which reduces its value if the dimension of the ordinal vectors is low. However, depending on the application, this fact can also be considered as an indicator of the existence of noise. 

A comparison with other classical metrics to evaluate synchronization has been carried out showing some similarities and differences. In general, $OS$ shows high correlation with $r$ and $PLV$, something that can be explained by the way $OS$ is constructed. Ordinal patterns filter part of the information contained in the amplitude of the signal, maintaining just the ranking in the time series. This is something between considering just the phase ($PLV$) or just the amplitude ($r$), since differences in amplitude are not related to changes in the $OS$ parameter as long as the ranking is not modified. 

In view of all, we believe that the use of $OS$ can be interesting (but not restricted to) for evaluating the amount of synchronization in neuroscientific data sets, where in-phase and anti-phase synchronization are know to co-exist, together with coordinations at different time scales.

 \section*{Appendix A: Coordination metrics}
 
\subsection*{A.1. Pearson's Correlation Coefficient} \label{PearsonCorrCoef} 

The Pearson's correlation coefficient $r$ consists of a covariance scaled by variances, thus capturing linear relationships among variables. From the equations of the variance (of $X$ and $Y$) and covariance (of $XY$), we obtain Pearson Correlation Coefficient as:

\begin{align}
S_{Y} &= \sqrt{\frac{\sum{(Y_{i}-\bar{Y})^2}}{n-1}} = \sqrt{\frac{\sum{y_{i}^2}}{n-1}} \\
S_{X} &= \sqrt{\frac{\sum{(X_{i}-\bar{X})^2}}{n-1}} = \sqrt{\frac{\sum{x_{i}^2}}{n-1}} \\
S_{XY} &= E[(X-E[X])(Y-E[Y])]
\\
r &= \frac{S_{XY}}{S_{X}{}S_{Y}} \label{R_XY}
\end{align}
Pearson's correlation is a measure of linear dependence between any pair of variables and it has the advantage of not requiring the knowledge of how variables are distributed. However, it should be applied only when variables are linearly related to each other.

\subsection*{A.2. Coherence}\label{Coherence}

Coherence (magnitude squared coherence or coherence spectrum) measures the linear correlation among the two spectra\citep{Pereda2005}. To calculate the coherence spectrum, data must be in the frequency domain. In order to do so, time series are usually divided into \textit{S} sections of equal size. The Fast Fourier Transform algorithm is then computed over the sections to get the estimate of each section's spectrum (periodogram). Then, the spectra of the sections is averaged to get the estimation of the whole data's spectrum (Welch's method). Finally, Coherence is a normalization of this estimate by the individual autospectral density functions \cite{Pereda2005}:

\begin{equation} \label{eqMSCohere}
 SC = \frac{|\langle Sp_{xy}^{2} \rangle|}{|\langle Sp_{xx} \rangle| |\langle Sp_{yy} \rangle|}
\end{equation}

where $Sp_{xy}$ is the Cross Power Spectral Density (CPSD) of both signals, $Sp_{xx}$ and $Sp_{yy}$ are the Power Spectral Density (PSD) of the segmented signals $X$ and $Y$ taken individually, and $\langle \cdot \rangle$ is the average over the S segments. In the case of the data sets obtained with the nonlinear electronic Lorenz systems, frequencies higher than $f_{cut}=7.5K$ Hz have been disregarded for the computation of $SC$, since the power spectra of the electronic circuits are completely flat above this frequency.
One of the drawbacks of Coherence is that  it doesn't discern the effects of amplitude and phase in the relationships measured between two signals, which makes its interpretation unclear \citep{Lachaux1999a,varela2001}.

\subsection*{A.3. Phase Locking Value}\label{PLV}
 
Phase Locking Value was first introduced by \citet{Lachaux1999a} as a new method to measure synchrony among neural populations. It has, at least, two major advantages over the classical coherence measure: it doesn't require data to be stationary, a condition that can rarely be validated; and has a relatively easy interpretation (in terms of phase coupling). However, the methods used to extract instantaneous phase, a step needed to calculate $PLV$ rely on stationarity, so indirectly $PLV$ can be affected by this condition \citep{Cohen2014}. To obtain the $PLV$, the signal has to be decomposed to it's instantaneous phases and amplitudes. To achieve this, there are several methods, such as Morlet wavelet convolution or Hilbert transform \citep{Pereda2005,Cohen2014}. In this work we will utilize the latter. Finally, $PLV$ is obtained averaging over time $t$: 

\begin{equation}
PLV = \frac{1}{N} \Abs{\sum_{n=1}^{N}{\exp{(i\theta(t,n))}}}
\label{eqPLV}
\end{equation}
where $\theta(t,n)$ is the (instantaneous) phase difference $\phi_{x} - \phi_{y}$, the phases to be compared from the signals $X$ and $Y$. Comparisons are carried out pairwise (bivariate). 

\subsection*{A.4. Mutual Information}\label{MI}

Mutual Information is a measure of shared information between any components of a system, between systems, or any other parameter whose value's probability can be estimated. It is based on Shannon's notion of entropy, which, in a general sense, tries to quantify the amount of information contained in a random variable by means of its estimated probability distribution. Mutual information measures the amount of information \textit{shared} between two random variables by means of its joint distribution, or conversely, the amount of information we can obtain from one random variable observing another. This is analogue to measuring the dependence between two random variables \citep{Veyrat-charvillon2009}. Let \textit{X} and \textit{Y} be two random variables with $\lbrace{x_{1},x_{2},...x_{n}}\rbrace$ and $\lbrace{y_{1},y_{2},...y_{n}}\rbrace$, $n$ possible values with probabilities $p(x)$ and $p(y)$. The $MI$ of \textit{X} relative to \textit{Y} can be written as:

\begin{equation}\label{eqMI}
MI({X}\cap{Y}) = \sum_{{x}\in{X}, \\ {y}\in{Y}}{p({x}\cap{y})}{}{\log_{2}}{}{\frac{p({x}\cap{y})}{p(x)p(y)}}
\end{equation}
\begin{equation}\label{eqMIcond1}
MI({X}\cap{Y}) = H(X) - H(X\vert{Y})
\end{equation}
where $p({x}\cap{y})$ is the probability that $X$ has a value of $x$ \textit{while} $Y$ has a value of $y$, $H(X)$ is the entropy of $X$ and $H({X}\vert{Y})$ is the conditional entropy of $X$ and $Y$.
One of the major advantages of $MI$ is that it captures linear and non-linear relationships among variables. One disadvantage is that it does not explicitly tell the shape of that distribution \citep{Cohen2014}. To get the mutual information between two random variables, we first need to estimate their probability density distribution \citep{Veyrat-charvillon2009,Cohen2014,Gierlichs2008}. 
Equation \ref{eqMI} compares joint probabilities against marginal ones. When two values are independent, the product of their marginal probabilities should equal their joint probability. When not, we can state that there is a relationship among them (not necessarily linear), because the probability of finding those values together is greater than the probability of finding them by chance. Thus, somehow, those time series are coupled, although we don't know the way it occurs.

\section*{Appendix B: Electronic version of the Lorenz system}
\label{S.6}

The equations of the master and slave electronic Lorenz systems are:

\begin{eqnarray}
V\dot{x}_{1} =\frac{1}{R_{1}C}\left(\frac{R_{1}}{R_{2}}V_{y_{1}} -\frac{R_{4}}{R_{3}}V_{x_{1}}+\frac{R_{4}}{R_{3}}V_{\xi_{1}}  \right) \\
V\dot{y}_{1} =\frac{1}{R_{5}C}\left(\frac{R_{5}}{R_{6}}V_{x_{1}}-\frac{R_{5}}{R_{7}}V_{x_{1}}V_{z_{1}}  \right)\\
V\dot{z}_{1} = \frac{1}{R_{8}C}\left(\frac{R_{8}}{R_{9}}V_{x_{1}}V_{y_{1}}-\frac{R_{11}}{R_{10}}V_{z_{1}} \right)\\
V\dot{x}_{2} =\frac{1}{R_{12}C}\left(\frac{R_{12}}{R_{13}}V_{y_{2}}-\frac{R_{15}}{R_{14}}V_{x_{2}} +\frac{R_{12}}{R_{29}}V_{\xi_{2}}+\frac{R_{12}}{R_{30}}V_{sinx} \right) \\
V\dot{y}_{2} =\frac{1}{R_{16}C}\left(\frac{R_{17}}{R_{16}}V_{x_{2}}-\frac{R_{17}}{R_{20}}V_{x_{2}}V_{z_{2}} \right)\\
V\dot{z}_{2} = \frac{1}{R_{19}C}\left(\frac{R_{19}}{R_{20}}V_{x_{2}}V_{y_{2}} -\frac{R_{22}}{R_{21}}V_{z_{2}} \right)
\end{eqnarray}
where $V_{x_{1,2}}$, $V_{y_{1,2}}$ and $V_{z_{1,2}}$ are the voltage variables of the master (sub-index 1) and slave (sub-index 2) Lorenz systems, $V_{in}=V_{x_1}-V_{x_2}$ is the coupling signal injected into the slave system in a diffusive way, $\kappa=\frac{R_{dp}}{C_5 R_{30}}$ is the coupling strength and $0 \leq R_{dp} \leq 1$ is the percentage of coupling controlled by the digital potentiometer. In the experiments where external noise
 is considered (see Appendix C), the amplitude of $V_{\xi_{1}}$ and $V_{\xi_{2}}$ are set to $0.5$ V and zero otherwise.

Table 1 contains the parameters of the resistances and capacitances used in the experiments. 

\begin{table}[h!]
\begin{tabular}{|l|l|l|}
\hline
$R_{1},R_{12}=100K\Omega$ & $R_{2},R_{13}=100K\Omega$ & $R_{3},R_{14}=10K\Omega$ \\
\hline
$R_{4},R_{15}=10K\Omega$  & $R_{5},R_{16}=1M\Omega$  & $R_{6},R_{17}=35.7K\Omega$ \\
\hline
$R_{7},R_{18}=20K\Omega$ & $R_{8},R_{19}=375K\Omega$ & $R_{9},R_{20}=20K\Omega$  \\
\hline
$R_{10},R_{21}=10K\Omega$ & $R_{11},R_{22}=10K\Omega$ & $R_{23}=10K\Omega$ \\
\hline
$R_{24}=10K\Omega$ & $R_{25}=10K\Omega$ & $R_{26}=10K\Omega$ \\
\hline
$R_{27}=10K\Omega$[0-1]  & $R_{28}=100K\Omega$ & $R_{29}=100K\Omega$ \\
\hline
$R_{30}=100K\Omega$ & $C_{1-6}=1nF$ & $V+=15V, V-=-15V$ \\
\hline
\end{tabular}
\caption{Parameters of the electronic components used for the construction of the Lorenz oscillators and the coupled circuit.}
\label{table:components}
\end{table}

\section*{Appendix C: Robustness of $OS$ in the presence of external noise}
\label{S.7}

\begin{figure}[h!]
	\begin{center}
		\includegraphics[width=0.45\linewidth,keepaspectratio]{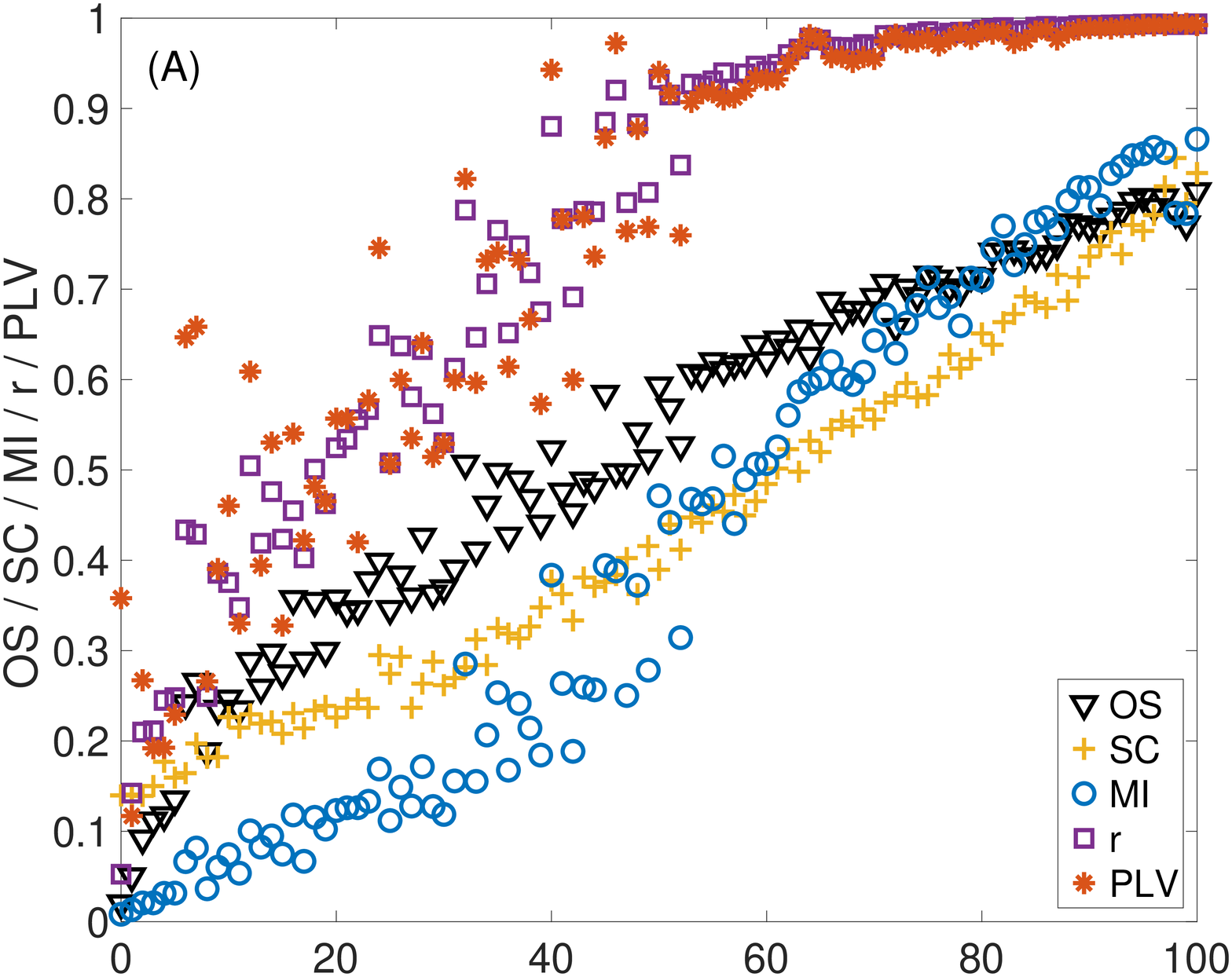}\hspace{0.5cm}
		\includegraphics[width=0.45\linewidth,keepaspectratio]{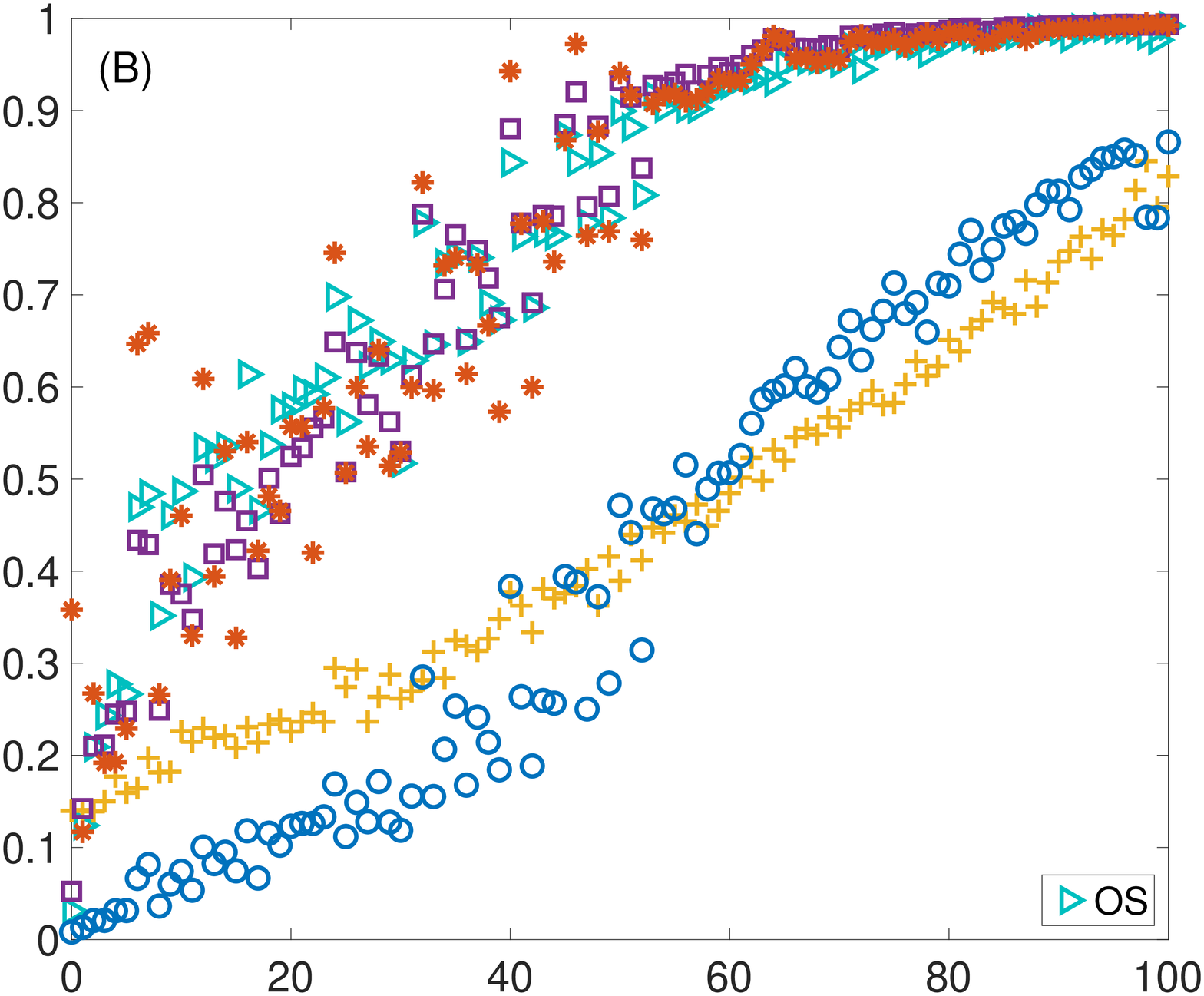}\vspace{0.35cm}
		\includegraphics[width=0.45\linewidth,keepaspectratio]{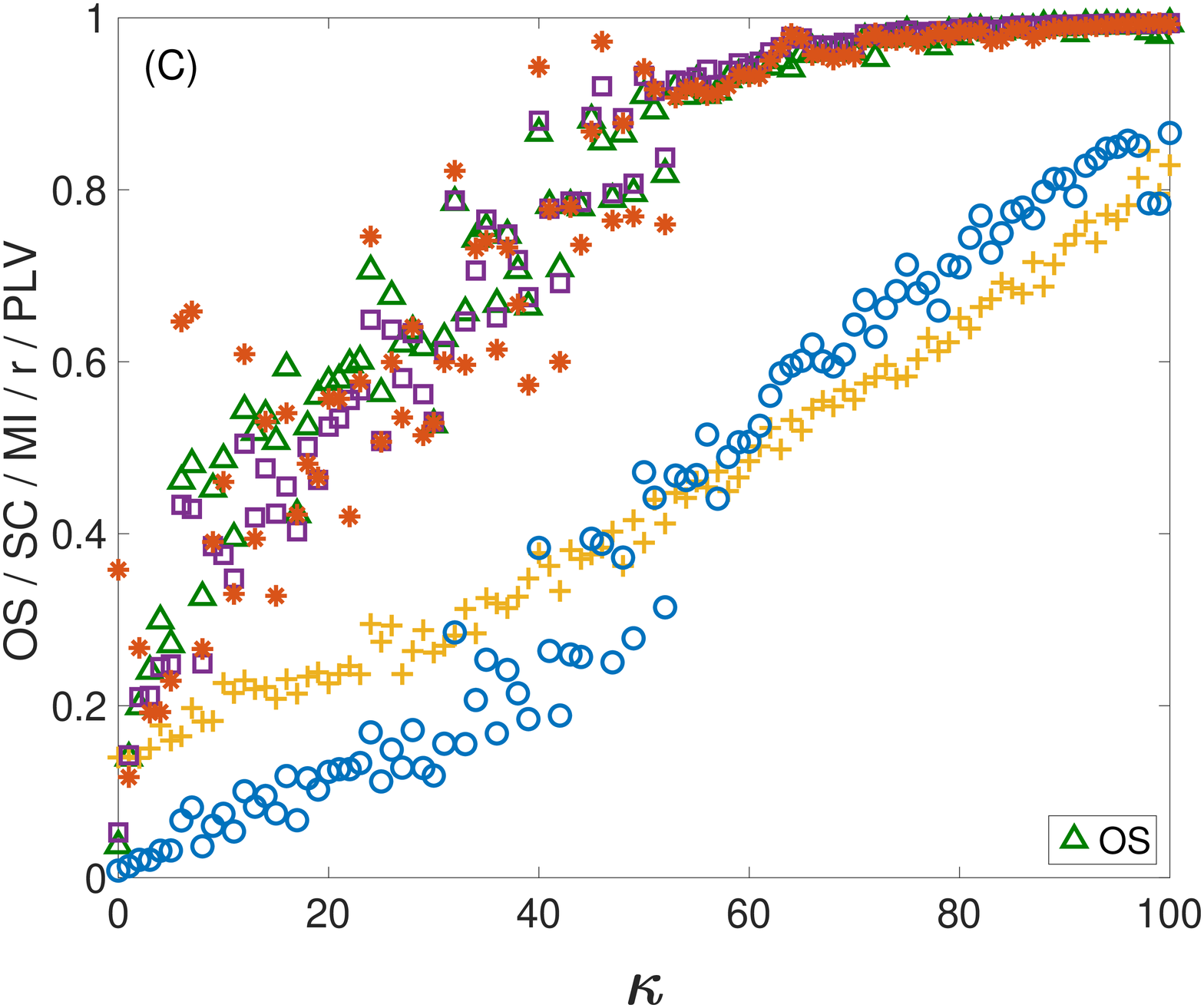}\hspace{0.5cm}
		\includegraphics[width=0.45\linewidth,keepaspectratio]{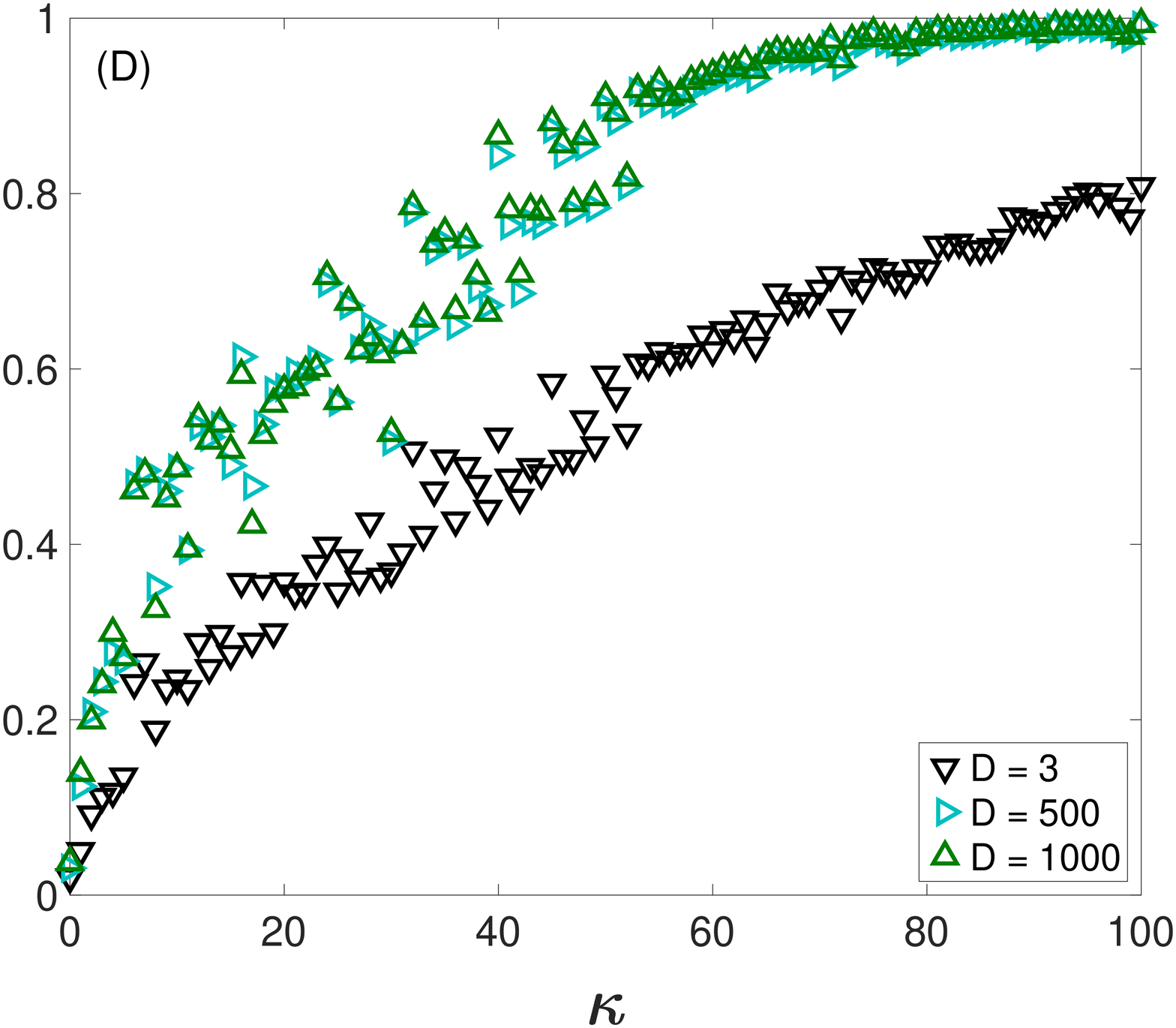}
		 \caption{Synchronization against coupling strength $\kappa$, as measured with mutual information $MI$ (light blue), spectral coherence $SC$ (yellow), phase locking value $PLV$ (red) Pearson correlation $r$ (purple) and ordinal synchronization $OS$ (black) for $D = 3$ (A), $D=500$ (B) and $D=1000$ (C). For comparison purposes, plot $D$ shows $OS$  against coupling strength for the different vector lengths, $D=3$ (black), $D=500$ (turquoise) and $D=1000$ (green).}\label{fig:fig08}
	\end{center}
\end{figure}

\begin{figure}[h]
	\begin{center}
		\includegraphics[width=0.45\linewidth,keepaspectratio]{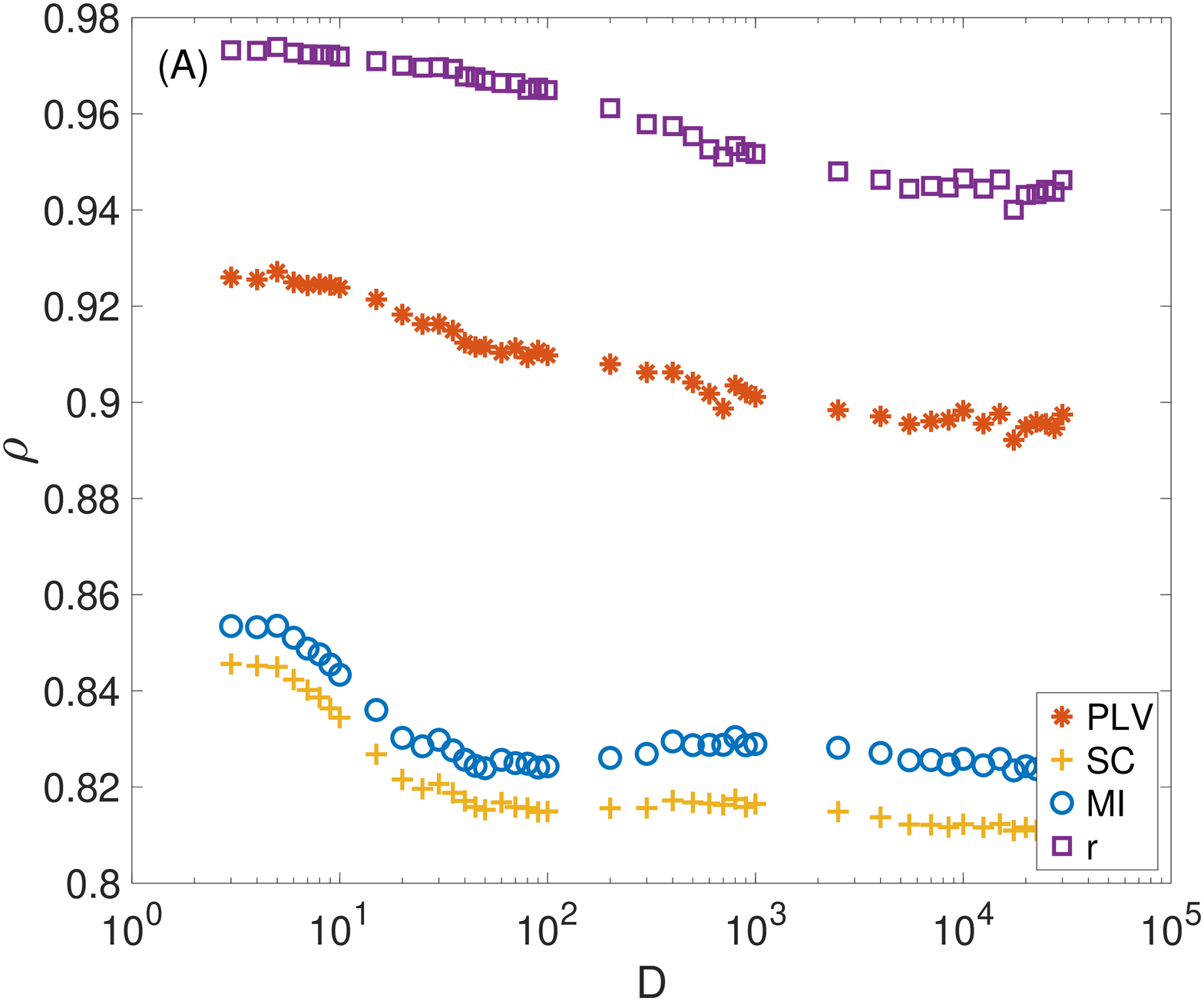}\hspace{0.5cm}
		\includegraphics[width=0.45\linewidth,keepaspectratio]{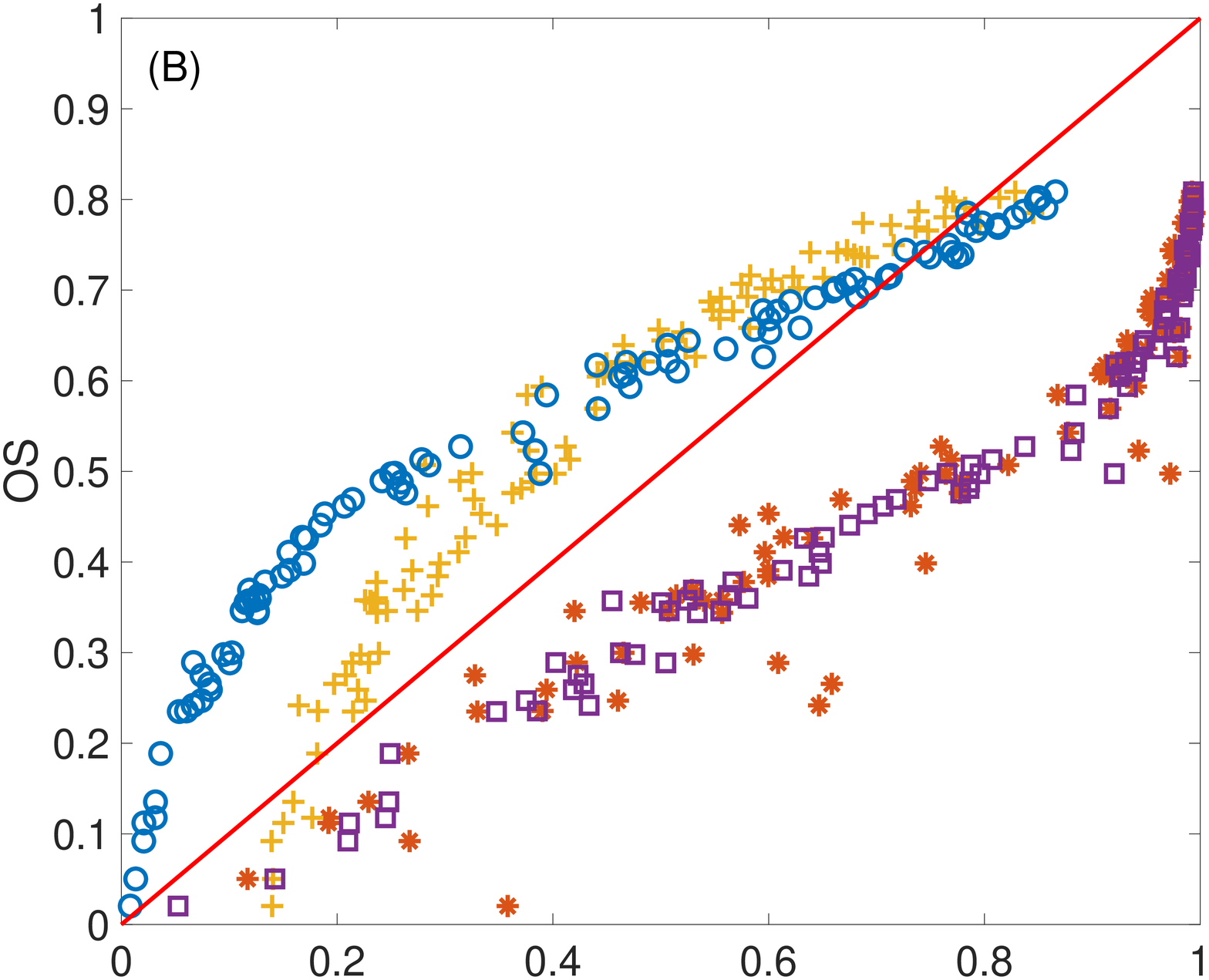}\vspace{0.35cm}
		\includegraphics[width=0.45\linewidth,keepaspectratio]{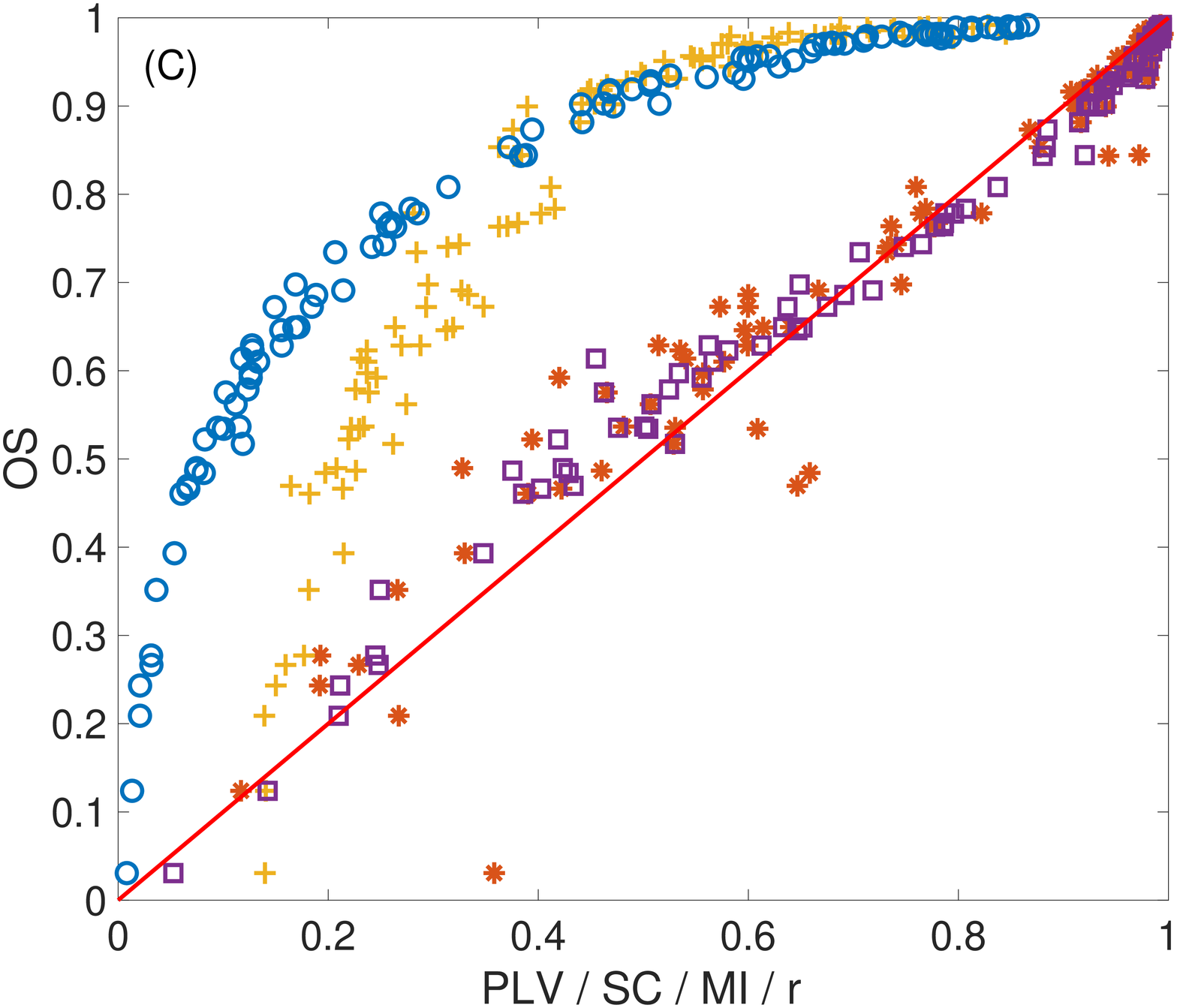}\hspace{0.5cm}
		\includegraphics[width=0.45\linewidth,keepaspectratio]{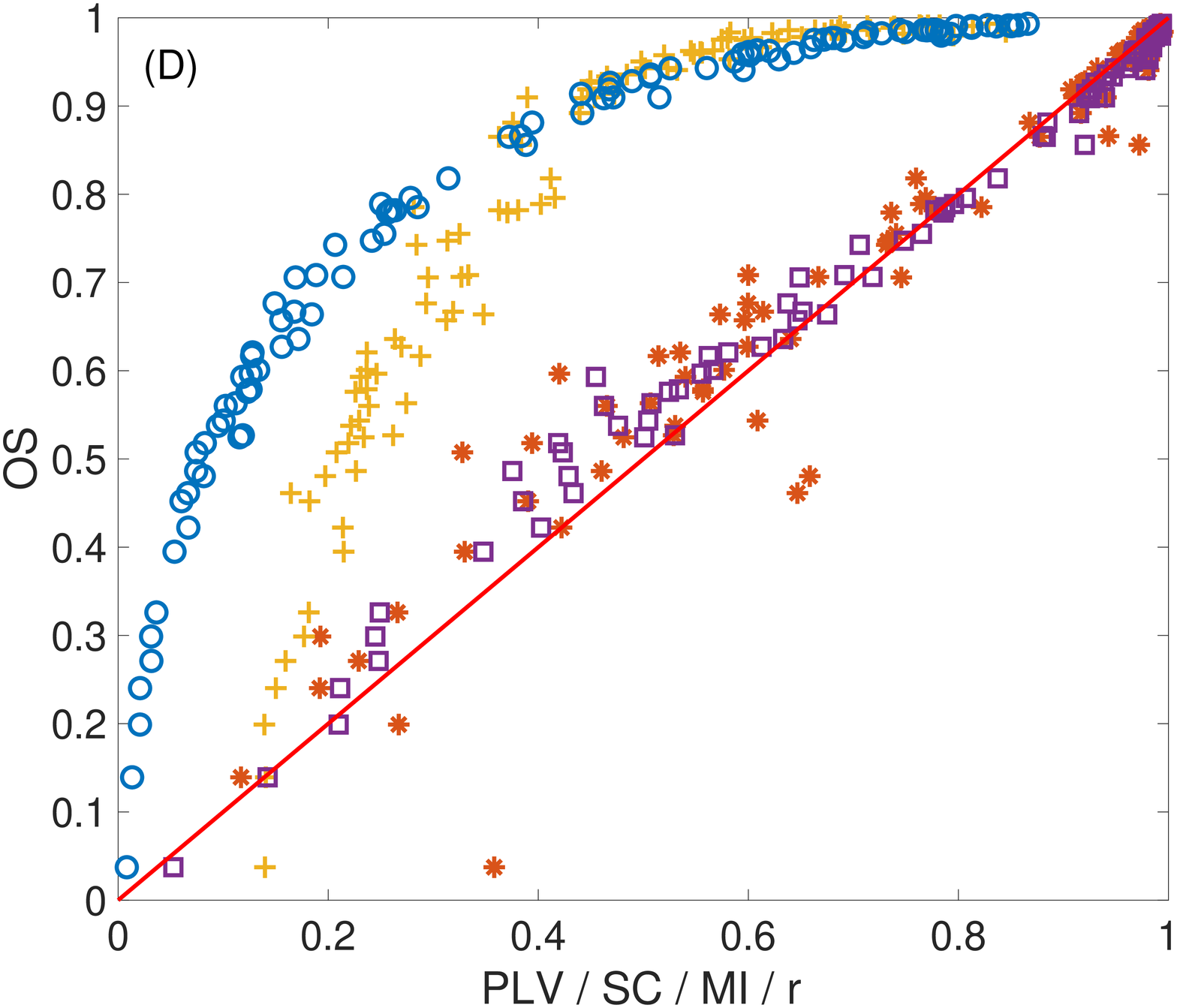}
		\caption{Panel A shows the average correlation between each synchronization measure and OS depending on the vector length ($D$) used to compute OS. Panels B-D show the correlation between OS and all other synchronization measures varying the coupling strength (from 0 to 100), for $D = 3$ (B), $D  = 500$ (C) and $D = 1000$ (D). Synchronization measures are Mutual information (MI; blue), Pearson correlation coefficient (r; purple), spectral coherence (SC; yellow) and phase locking value (PLV; red). The red line corresponds to $y=x$.}\label{fig:fig09}
	\end{center}
\end{figure}

Figures \ref{fig:fig08}-\ref{fig:fig09} are equivalent to Figs.  \ref{fig:fig03}-\ref{fig:fig04} but in the presence of external noise. In this case, we have introduced two noises $\xi_1$ and $\xi_{2}$ perturbing the $x_1$ and $x_2$ variables of the master and slave Lorenz systems as explained in Appendix B. Comparing Fig. \ref{fig:fig08} and Fig. \ref{fig:fig03} we can observe that all synchronization metrics have reduced their values in the presence of external noise, however, the behaviour remains qualitatively similar to the one reported in Fig. \ref{fig:fig03}. Again, the case $D=3$ is the one suffering the most from the presence of noisy signals (Fig. \ref{fig:fig08}D). 
When comparing $OS$ with the rest of synchronization metrics (Fig. \ref{fig:fig09}), we can also observe a reduction of the correlations respect to the case without external noise. Again, $r$ and $PLV$ are the metrics showing higher correlation with $OS$, having a linear correlation for $D=500$ and $D=1000$. This correlation is impaired for $D=3$, since it corresponds to the ordinal vector length that is more affected by noise. On the other hand, the nonlinear correlations with $MI$ and $SC$ remain quite similar as in the case of the absence of external noise.

\clearpage

\section*{References}
\label{S.8}


\bibliographystyle{model1-num-names}
\bibliography{OS.bib}







\end{document}